\documentclass[a4paper,12pt]{article}

\pdfoutput=1

\usepackage{amsmath}
\usepackage{amssymb}
\usepackage{amsfonts}
\usepackage{mathrsfs}
\usepackage{bbm}
\usepackage{graphicx,subfigure,booktabs,adjustbox}
\usepackage[numbers,sort&compress]{natbib}
\usepackage{verbatim}

\usepackage[colorlinks,
            linkcolor=black,
            filecolor=black,
            anchorcolor=black,
            urlcolor=black,
            citecolor=black
            ]{hyperref}

\usepackage{color}
\usepackage{ulem}
\usepackage{setspace}
\numberwithin{equation}{section}

\newlength{\dinwidth}
\newlength{\dinmargin}
\makeatletter
\newcommand{\thickhline}{%
    \noalign {\ifnum 0=`}\fi \hrule height 1pt
    \futurelet \reserved@a \@xhline
}
\makeatother

\allowdisplaybreaks

\begin{document}

\title{\bf \Large Revisiting the one leptoquark solution to the $R(D^{(\ast)})$ anomalies and its phenomenological implications}

\author{
Xin-Qiang Li\footnote{xqli@mail.ccnu.edu.cn},
Ya-Dong Yang\footnote{yangyd@mail.ccnu.edu.cn}\,
and
Xin Zhang\footnote{zhangxin027@mails.ccnu.edu.cn}\\[15pt]
\small Institute of Particle Physics and Key Laboratory of Quark and Lepton Physics~(MOE), \\
\small Central China Normal University, Wuhan, Hubei 430079, China}

\date{}
\maketitle
\vspace{0.2cm}

\begin{abstract}
{\noindent}It has been shown recently that the anomalies observed in $\bar B\to D^{(\ast)}\tau\bar\nu_\tau$ and $\bar B\to \bar K\ell^+\ell^-$ decays could be resolved with just one scalar leptoquark. Fitting to the current data on $R(D^{(\ast)})$ along with acceptable $q^2$ distributions in $\bar B\to D^{(\ast)}\tau\bar\nu_\tau$ decays, four best-fit solutions for the operator coefficients have been found. In this paper, we explore the possibilities of how to discriminate these four solutions. Firstly, we find that two of them are already excluded by the decay $B_c^-\to \tau^-\bar\nu_\tau$, because the predicted decay widths have already overshot the total width $\Gamma_{B_c}$. It is then found that the remaining two solutions result in two effective Hamiltonians governing $b\to c \tau \bar\nu_{\tau}$ transition, which differ by a sign and enhance the absolute value of the coefficient of $\bar c_L\gamma_\mu b_L\,\bar \tau_L\gamma^\mu{\nu_\tau}_L$ operator by about $12\%$. However, they give nearly the same predictions as in the SM for the  $D^\ast$ and $\tau$ longitudinal polarizations as well as the lepton forward-backward asymmetries in $\bar B\to D^{(\ast)}\tau\bar\nu_\tau$ decays. For the other observables like $\mathcal B(B_c^-\to \tau^-\bar\nu_\tau)$, $\mathcal B(B_c^- \to \gamma\tau^-\bar\nu_\tau)$, $R_{D^{(\ast)}}(q^2)$, ${\rm d}\mathcal B(\bar B\to D^{(\ast)}\tau\bar\nu_\tau)/{\rm d}q^2$ and $\mathcal B(\bar B\to X_c\tau\bar\nu_\tau)$, on the other hand, the two solutions give sizable enhancements relative to the SM predictions. With measurement of $B_c^-\to \tau^-\bar\nu_\tau$ at LHCb and refined measurements of observables in $\bar B\to D^{(\ast)}\tau\bar\nu_\tau$ at both LHCb and Belle-II, such a specific NP scenario could be further deciphered.
\end{abstract}

\newpage

\section{Introduction}
\label{sec:intro}

With the discovery of heavy quark spin-flavor symmetry and the formulation of heavy quark effective theory~(HQET)~\cite{Grinstein:1990mj,Eichten:1989zv,Georgi:1990um,Manohar:2000dt,Neubert:1993mb}, it has become clear that the physical observables in semi-leptonic $\bar B\to D^{(\ast)}\ell\bar\nu_\ell$ could be rather reliably predicted within the Standard Model~(SM), especially at the zero recoil point, allowing therefore a reliable determination of the Cabibbo-Kobayashi-Maskawa~(CKM) element $V_{cb}$. It is also believed that the effect of New Physics~(NP) beyond the SM should be tiny since these decays are induced by the tree-level charged current.

However, the BaBar~\cite{Lees:2012xj,Lees:2013uzd}, Belle~\cite{Huschle:2015rga,Abdesselam:2016cgx} and LHCb~\cite{Aaij:2015yra} collaborations have recently observed anomalies in the ratios
\begin{equation}
R(D^{(\ast)})=\frac{{\mathcal B}(\bar B\to D^{(\ast)}\tau\bar\nu_\tau)}
{{\mathcal B}(\bar B\to D^{(\ast)}\ell\bar\nu_\ell)}\,,\qquad \ell=e/\mu\,.
\end{equation}
The Heavy Flavor Average Group~(HFAG) gives the average values~\cite{HFAG:2016}
\begin{align}\label{eq:experiment}
R(D)_{\rm avg}=&0.397\pm0.040\pm0.028\,,\nonumber\\
R(D^\ast)_{\rm avg}=&0.316\pm0.016\pm0.010\,,
\end{align}
which exceed the SM predictions~\cite{Na:2015kha,Fajfer:2012vx}
\begin{align}\label{eq:RD}
R(D)_{\rm SM}=&0.300\pm 0.008\,,\nonumber\\
R(D^\ast)_{\rm SM}=&0.252 \pm 0.003\,,
\end{align}
by $1.9\sigma$ and $3.3\sigma$, respectively. Especially when the $R(D)$-$R(D^{\ast})$ correlation of $-0.21$ is taken into account, the tension with the SM predictions would be at $4.0\sigma$ level~\cite{HFAG:2016}. Theoretically, $R(D)$ and $R(D^{\ast})$ can be rather reliably calculated, because they are independent of the CKM element $|V_{cb}|$ and, to a large extent, of the $B\to D^{(\ast)}$ transition form factors.

The above anomalies have been investigated extensively both within model-independent frameworks~\cite{Freytsis:2015qca,Calibbi:2015kma,Alonso:2015sja,Tanaka:2012nw,
Fajfer:2012jt,Becirevic:2016hea,Bhattacharya:2015ida,Bhattacharya:2014wla,Duraisamy:2014sna,
Hagiwara:2014tsa,Dutta:2013qaa,Duraisamy:2013kcw,Biancofiore:2013ki,Bailey:2012jg,
Becirevic:2012jf,Datta:2012qk,Faller:2011nj,Chen:2005gr,Fan:2013qz,Fan:2015kna,Alok:2016qyh,Ivanov:2016qtw}, as well as in some specific NP models where the $b\to c\tau\bar{\nu}_{\tau}$ transition is mediated by leptoquarks~\cite{Deppisch:2016qqd,Dumont:2016xpj,Dorsner:2016wpm,Freytsis:2015qca,Calibbi:2015kma,Sakaki:2014sea,Bauer:2015knc,
Fajfer:2015ycq,Sahoo:2015pzk,Barbieri:2015yvd,Sakaki:2013bfa}, charged Higgses~\cite{Freytsis:2015qca,Celis:2012dk,Cline:2015lqp,Kim:2015zla,Crivellin:2015hha,
Hwang:2015ica,Crivellin:2013wna,Sakaki:2012ft,Nierste:2008qe,Kiers:1997zt,Tanaka:1994ay,
Hou:1992sy}, charged vector bosons~\cite{Boucenna:2016wpr,Hati:2015awg,Greljo:2015mma,Freytsis:2015qca}, and sparticles~\cite{Das:2016vkr,Zhu:2016xdg,Deshpande:2012rr}. It is also interesting to point out that, besides the branching ratios, the measured differential distributions $d\Gamma(\bar B \to D^{(\ast)}\tau\bar{\nu})/dq^2$ by BaBar~\cite{Lees:2013uzd} and Belle~\cite{Huschle:2015rga,Abdesselam:2016cgx} provide very complementary information to distinguish NP from the SM as well as different NP models from each other~\cite{Freytsis:2015qca,Sakaki:2014sea,Celis:2012dk}.

With both the ratios $R(D^{(\ast)})$ and the $q^2$ spectra taken into account, Freytsis, Ligeti and Ruderman have identified viable models with leptoquark mediators, which are consistent with minimal flavor violation and could provide good fits to the current data; especially, four best-fit solutions are found for the operator coefficients induced by scalar leptoquarks~\cite{Freytsis:2015qca}. With this observation, Bauer and Neubert have recently proposed a very simple NP model by extending the SM with a single TeV-scale scalar leptoquark $\phi$ transforming as $(\mathbf{3}, \mathbf{1},-\frac{1}{3})$ under the SM gauge group, and shown that the anomalies observed in $\bar B \to D^{(\ast)}\tau\bar{\nu}$, $\bar B\to \bar K\ell^+\ell^-$~\cite{Aaij:2014ora}, as well as the anomalous magnetic moment of muon~\cite{Davier:2010nc} can be explained in a natural way, while constraints from other precision measurements in the flavor sector are also satisfied without fine-tuning~\cite{Bauer:2015knc}.

To further test such an interesting scenario, in this paper, we shall explore in detail the effect of the scalar leptoquark on the purely leptonic $B_c^-\to\tau^-\bar\nu_\tau$, the radiative leptonic $B_c^-\to\gamma\tau^-\bar\nu_\tau$, the exclusive semi-leptonic $\bar B\to D^{(\ast)}\tau\bar{\nu}_{\tau}$, and the inclusive semi-leptonic $B\to X_c\tau\bar{\nu}_{\tau}$ decays. It is found that two of the four best-fit solutions obtained in Ref.~\cite{Freytsis:2015qca} are already excluded by the decay $B_c^-\to \tau^-\bar\nu_\tau$, because the predicted decay widths have already overshot the total width $\Gamma_{B_c}$. The remaining two solutions result in two effective Hamiltonians that differ by a sign, but give almost the same predictions as in the SM for the $D^\ast$ and $\tau$ longitudinal polarizations as well as the lepton forward-backward asymmetries in $\bar B\to D^{(\ast)}\tau\bar\nu_\tau$ decays. For the observables $\mathcal B(B_c^-\to \tau^-\bar\nu_\tau)$, $\mathcal B(B_c^- \to \gamma\tau^-\bar\nu_\tau)$, $R_{D^{(\ast)}}(q^2)$, ${\rm d}\mathcal B(\bar B\to D^{(\ast)}\tau\bar\nu_\tau)/{\rm d}q^2$ and $\mathcal B(\bar B\to X_c\tau\bar\nu_\tau)$, on the other hand, the two solutions give sizable enhancements relative to the SM predictions. With measurement of $B_c^-\to \tau^-\bar\nu_\tau$ at LHCb and refined measurements of observables in $\bar B\to D^{(\ast)}\tau\bar\nu_\tau$ at both LHCb and Belle-II, such a specific NP scenario could be further deciphered.

This paper is organized as follows: In section~\ref{sec:framework}, we recapitulate the scenario with just one scalar leptoquark introduced in Ref.~\cite{Bauer:2015knc}. In section~\ref{sec:analytic calculation}, we consider the purely leptonic $B_c^-\to\tau^-\bar\nu_\tau$ decay, from which two of the four best-fit solutions for the operator coefficients are found to be already excluded. The effects of the remaining two solutions on $B_c^-\to\gamma\tau^-\bar\nu_\tau$, $\bar B\to D^{(\ast)}\tau\bar\nu_\tau$ and $B\to X_c\tau\bar\nu_\tau$ decays are then investigated in section~\ref{sec:numerical analysis}. Our conclusions are finally made in section~\ref{sec:conclusion}. Appendixes~A and B contain the formulae relevant to these decays.

\section{The one scalar leptoquark scenario}
\label{sec:framework}

In this section, we recapitulate the model proposed very recently by Bauer and Neubert~\cite{Bauer:2015knc}, where a single TeV-scale leptoquark $\phi$ is added to the SM to address the aforementioned anomalies in flavor physics. The new scalar $\phi$ transforms as $(\mathbf{3}, \mathbf{1}, -\frac{1}{3})$ under the SM gauge group, and its couplings to fermions are described by the Lagrangian~\cite{Dorsner:2016wpm,Bauer:2015knc}
\begin{equation}
\mathcal L_{\rm int}^\phi=\bar Q_L^c{\boldsymbol\lambda}^Li\tau_2L\phi^{\ast}+\bar u_R^c{\boldsymbol\lambda}^Rl_R\phi^{\ast}+{\rm h.c.}\,,
\end{equation}
where ${\boldsymbol\lambda}^{L,R}$ are the Yukawa coupling matrices in flavor space, $Q_L,\,L$ denote the left-handed quark and lepton doublet, $u_R,\,l_R$ the right-handed up-type quark and lepton singlet respectively, and $\psi^c=C\bar{\psi}^T$, $\bar{\psi}^c=\psi^T C$~($C=i\gamma^2\gamma^0$) are the charge-conjugated spinors. Rotating the Lagrangian from the weak to the mass basis for quarks and charged leptons, the interaction terms take the form
\begin{equation}\label{eq:Lint}
\mathcal L_{\rm int}^\phi=\bar u_L^c{\boldsymbol\lambda}_{ul}^Ll_L\phi^{\ast}-\bar d_L^c{\boldsymbol\lambda}_{d\nu}^L\nu_L\phi^{\ast}+\bar u_R^c{\boldsymbol\lambda}_{ul}^Rl_R\phi^{\ast}+{\rm h.c.}\,,
\end{equation}
where ${\boldsymbol\lambda}_{ul}^L$, ${\boldsymbol\lambda}_{d\nu}^L$ and ${\boldsymbol\lambda}_{ul}^R$ are now the coupling matrices in mass basis, and describe the strength of $\phi$ interactions with fermions.

Writing down the tree-level $\phi$-exchange amplitude for the process $b\to c\tau\bar{\nu_{\tau}}$ in the leading order in $k^2/M_{\phi}^2$ expansion, where $k\sim \mathcal{O}(m_b)$ is the momentum flowing through the $\phi$ propagator and $M_{\phi}$ the leptoquark mass, and then performing the Fierz transformation of the resulting four-fermion operators, one can get the effective Hamiltonian
\begin{align}\label{eq:phi Hamiltonian}
\mathcal H_{\rm{eff}}^\phi=&-\frac{1}{2M_{\phi}^2}\Big[\lambda^L_{b\nu_{\tau}}\lambda_{c\tau}^{L*}\bar\, b^c\gamma_{\mu}P_Rc^c\,\bar{\tau}\gamma^{\mu}P_L\nu_{\tau}+\lambda^L_{b\nu_{\tau}}\lambda_{c\tau}^{R*}\, (\bar b^cP_Lc^c\,\bar{\tau}P_L\nu_{\tau}+\frac{1}{4}\bar b^c\sigma_{\mu\nu}P_Lc^c\,\bar{\tau}\sigma^{\mu\nu}P_L\nu_{\tau})\Big],
\end{align}
where $P_{L,R} = (1 \mp \gamma_5)/2$ are the chirality projectors. Using the definitions $\psi^c=C\bar{\psi}^T$, $\bar{\psi}^c=\psi^T C$, and the relations $C\gamma_{\mu}=-\gamma_{\mu}^T C$, $C\gamma_5=\gamma_5^T C$, one can easily arrive at the equations
\begin{align}
\bar b^c\gamma_{\mu}P_Rc^c=&-\bar c\gamma_{\mu}P_Lb\,,\\
\bar b^cP_Lc^c=&\;\bar cP_Lb\,,\\
\bar b^c\sigma_{\mu\nu}P_Lc^c=&-\bar c\sigma_{\mu\nu}P_Lb\,.
\end{align}
Plugging the above three equations into Eq.~(\ref{eq:phi Hamiltonian}) and including also the SM contribution, one obtains then the total effective Hamiltonian governing the $b\to c\tau\bar{\nu_{\tau}}$ transition
\begin{align}\label{eq:total Hamiltonian}
\mathcal H_{\text{eff}}=&\frac{4G_F}{\sqrt2}V_{cb}\left[C_V(M_\phi)\,\bar c\gamma_\mu P_L b\,\bar\tau\gamma^\mu P_L\nu_\tau+C_S(M_\phi)\,\bar c P_L b\,\bar\tau P_L\nu_\tau-\frac{1}{4}C_T(M_\phi)\,
\bar c\sigma_{\mu\nu}P_Lb\,\bar{\tau}\sigma^{\mu\nu}P_L\nu_{\tau}\right],
\end{align}
where $C_V$, $C_S$, $C_T$ are the Wilson coefficients of the corresponding operators at the matching scale $\mu=M_\phi$, and are given explicitly as
\begin{align}\label{eq:WCmuphi}
C_V(M_\phi)=&1+\frac{\lambda_{b\nu_{\tau}}^L\lambda_{c\tau}^{L\ast}}
{4\sqrt{2}G_FV_{cb}M_{\phi}^2}\nonumber\,,\\
C_S(M_\phi)=C_T(M_\phi)=&-\frac{\lambda_{b\nu_{\tau}}^L\lambda_{c\tau}^{R\ast}}
{4\sqrt{2}G_FV_{cb}M_{\phi}^2}\,.
\end{align}

In order to re-sum potentially large logarithmic effects and to make predictions for physical observables, the Wilson coefficients given by Eq.~(\ref{eq:WCmuphi}) should be run down to the characteristic scale of the processes we are interested in, i.e., $\mu_b\sim m_b$. While the vector current is conserved and needs not be renormalized, the evolutions at the leading logarithmic approximation of the scalar $C_S$ and tensor $C_T$ coefficients are given, respectively, by~\cite{Dorsner:2013tla}
 \begin{align}\label{eq:WCmub}
      C_S(\mu_b) =&\left[ \alpha_s(m_t) \over \alpha_s(\mu_b) \right]^{\gamma_S \over 2\beta_0^{(5)}} \left[ \alpha_s(M_\phi) \over \alpha_s(m_t) \right]^{\gamma_S \over 2\beta_0^{(6)}} C_S(M_{\phi}) \,,\nonumber \\
      C_T(\mu_b) =&\left[ \alpha_s(m_t) \over \alpha_s(\mu_b) \right]^{\gamma_T \over 2\beta_0^{(5)}} \left[ \alpha_s(M_\phi) \over \alpha_s(m_t) \right]^{\gamma_T \over 2\beta_0^{(6)}} C_T(M_{\phi})\,,
\end{align}
where $\gamma_S=-8$~\cite{Chetyrkin:1997dh} and $\gamma_T=8/3$~\cite{Gracey:2000am} are the LO anomalous dimensions of QCD scalar and tensor currents respectively, and $\beta_0^{(f)}=11-2n_f/3$ the LO beta function coefficient, with $n_f$ being the number of active quark flavors.

As detailed in Ref.~\cite{Freytsis:2015qca}, such a scalar leptoquark scenario introduced above could provide good explanations to the $R(D)$ and $R(D^\ast)$ anomalies along with acceptable $q^2$ spectra. Taking $M_{\phi}=1~{\rm TeV}$ as a benchmark and performing a two-dimensional $\chi^2$ fit, they found four best-fit solutions with $\chi_{\rm min}^2<5$ for the operator coefficients, which are listed below and denoted, respectively, as $P_A$, $P_B$, $P_C$, $P_D$:
\begin{align}\label{eq:parameter}
(\lambda_{b\nu_{\tau}}^L\lambda_{c\tau}^{L\ast},
 \lambda_{b\nu_{\tau}}^L\lambda_{c\tau}^{R\ast}) =
(C_{S_R}^{\prime\prime}, C_{S_L}^{\prime\prime}) = &
\left\{
      \begin{array}{lll}
        (\phantom{-}0.35,  &           -0.03),  & P_A\\
        (\phantom{-}0.96,  & \phantom{-}2.41),  & P_B\\
        (          -5.74,  & \phantom{-}0.03),  & P_C\\
        (          -6.34,  &           -2.39),  & P_D
      \end{array}
    \right.\,.
\end{align}
It is noticed that only the solution $P_A$ is adopted by Bauer and Neubert~\cite{Bauer:2015knc}, arguing that the other three require significantly larger couplings. It would be worth investigating whether the four best-fit solutions could be discriminated from each other using the processes mediated by the same effective operators given by Eq.~(\ref{eq:total Hamiltonian}). To this end, in addition to $\bar B\to D^{(\ast)}\tau\bar\nu_\tau$, we shall examine their effects on $B_c^-\to\tau^-\bar\nu_\tau$, $B_c^-\to\gamma\tau^-\bar\nu_\tau$ and $B\to X_c\tau\bar\nu_\tau$ decays.

As a final comment, it should be noted that the interaction Lagrangian Eq.~\eqref{eq:Lint} also gives rise to tree-level neutral quark and lepton currents; after integrating out the scalar leptoquark and performing the Fierz transformation, one encounters the operators $(\bar{u}_{i}u_{j})(\ell^{+}\ell^{-})$ and $(\bar{d}_{i}d_{j})(\bar{\nu}\nu)$, the Wilson coefficients of which can be constrained, for example, by the rare decays $D^0\to \mu^+\mu^-$ and $D^+\to \pi^+\mu^+\mu^-$, as well as $B\to X_s\nu\bar\nu$, $B\to K^{(\ast)}\nu\bar\nu$ and $K\to \pi\nu\bar\nu$, respectively. For more information about the low-energy constraints on the model, we refer the reader to Refs.~\cite{Freytsis:2015qca,Dorsner:2016wpm,Bauer:2015knc}.

\section{The effects of scalar leptoquark in $B_c$ and $B$ decays}
\label{sec:analytic calculation}

In this section, we explore the effects of the scalar leptoquark $\phi$ with the four best-fit solutions in the $B_c$ and $B$-meson decays.

\subsection{Purely leptonic decay $B_c^-\to\tau^-\bar\nu_\tau$}

Firstly, we investigate the effect of $\phi$ on the purely leptonic decay $B_c^-\to\tau^-\bar\nu_\tau$, the decay amplitude of which, including both the SM and NP contributions, can be written as
\begin{align}
\mathcal A(B_c^-\to\tau^-\bar\nu_\tau) =& i\frac{G_FV_{cb}}{\sqrt{2}}\Big[C_V\,\langle 0|\bar c\gamma_\mu\gamma_5b|B_c\rangle\,\bar\tau\gamma^\mu(1-\gamma_5)\nu_\tau + C_S\,\langle 0|\bar c\gamma_5b|B_c\rangle\,\bar\tau(1-\gamma_5)\nu_\tau\Big]\,.
\end{align}
Together with the definition of the $B_c$-meson decay constant $f_{B_c}$, $\langle 0|\bar c\gamma^\mu\gamma_5b|B_c(p)\rangle = if_{B_c}p^{\mu}$, and using the equation of motion, one can express the matrix element of pseudoscalar current as $\langle 0|\bar c\gamma_5b|B_c(p)\rangle = -\frac{if_{B_c}m_{B_c}^2}{m_b(\mu_b)+m_c(\mu_b)}$. The decay width for this process then reads
\begin{align}\label{eq:Bcwidth}
\Gamma(B_c^-\to\tau^-\bar\nu_\tau)={G_F^2\over 8\pi}|V_{cb}|^2f_{B_c}^2m_{B_c}^3\frac{m_\tau^2}{m_{B_c}^2}\left(1-{m_\tau^2\over m_{B_c}^2}\right)^2\,\left|C_V-C_S{m_{B_c}^2\over m_\tau\big[m_b(\mu_b)+m_c(\mu_b)\big]}\right|^2\,,
\end{align}
where $C_V$ and $C_S$ are the Wilson coefficients of (axial)vector and (pseudo)scalar operators, and $m_b$ and $m_c$ the current quark masses, all being given at the scale $\mu_b=m_b$.

With the input parameters collected in Table~\ref{tab:inputs} and $f_{B_c}=0.434~{\rm GeV}$~\cite{Colquhoun:2015oha}, we get
\begin{align}\label{eq:Bc result}
{\rm \Gamma}(B_c^-\to\tau^-\bar\nu_\tau)=&\left\{
            \begin{array}{rr}
              2.22\times 10^{-2}\,\Gamma_{B_c}, & {\rm SM} \\
              2.45\times 10^{-2}\,\Gamma_{B_c}, & P_A\\
              1.33\,\Gamma_{B_c},  & P_B\\
              2.39\times 10^{-2}\,\Gamma_{B_c}, & P_C \\
              1.31\,\Gamma_{B_c},  & P_D
            \end{array}
          \right.\,,
\end{align}
which are normalized to the total decay width $\Gamma_{B_c}=1/\tau_{B_c}=1/0.507~{\rm ps}^{-1}$~\cite{Agashe:2014kda}. The results labelled by $P_{A,B,C,D}$ are obtained using the four best-fit solutions given by Eq.~(\ref{eq:parameter}) and with $M_{\phi}=1~{\rm TeV}$. Clearly, one can see that two of the four solutions, $P_B$ and $P_D$, are already excluded by the decay $B_c^-\to \tau^-\bar\nu_\tau$, because the predicted decay widths have already overshot the total width $\Gamma_{B_c}$. Therefore, in the following, we need only consider the remaining two best-fit solutions $P_A$ and $P_C$.

\subsection{Comparison between the solutions $P_A$ and $P_C$}

Before going to detail their effects on the other decays, we give firstly a comparison between the remaining two best-fit solutions $P_A$ and $P_C$. Plugging into the effective Hamiltonian Eq.~\eqref{eq:total Hamiltonian} the fitted values of the effective couplings in $P_A$ and $P_C$ solutions, we get
\begin{align}\label{eq:hefffit}
\mathcal H_{\text{fit}} = &\frac{4G_F}{\sqrt2}V_{cb}\left\{
\Bigg[1+\Bigg(
\begin{array}{rr}
\phantom{-}0.129 & \text{for $P_A$} \\
          -2.117 & \text{for $P_C$} \\
\end{array}
\Bigg)
\Bigg]\bar c\gamma_\mu P_L b\,\bar\tau\gamma^\mu P_L\nu_\tau+
\Bigg(
\begin{array}{rr}
\phantom{-}0.018 & \text{for $P_A$} \\
          -0.018 & \text{for $P_C$} \\
\end{array}
\Bigg)\bar c P_L b\,\bar\tau P_L\nu_\tau\right.\nonumber\\
&\left. \hspace{1.8cm} +\Bigg(
\begin{array}{rr}
          -0.002 & \text{for $P_A$} \\
\phantom{-}0.002 & \text{for $P_C$} \\
\end{array}
\Bigg)\bar c\sigma_{\mu\nu}P_Lb
\,\bar{\tau}\sigma^{\mu\nu}P_L\nu_{\tau}\right\}\,,
\end{align}
which have been run down from the scale $\mu=M_\phi=1~{\rm TeV}$ to the scale $\mu=m_b=4.18~{\rm GeV}$.

It is observed that the coefficients of the operator $\bar c\gamma_\mu P_L b\,\bar\tau\gamma^\mu P_L\nu_\tau$,
\begin{equation}
C_V^{\text{fit}}=\left\{\begin{array}{rr}
                         \phantom{-}1.129, & \text{for $P_A$} \\
                                   -1.117, & \text{for $P_C$}
                        \end{array}
\right.\,,
\end{equation}
have nearly the same absolute values, both enhancing the SM result by $\sim12\%$, but the sign of solution $P_C$ is flipped relative to the SM. Furthermore, the $\sim1.2\%$ difference in $|C_V^{\text{fit}}|$ would be too small to be discriminated from each other phenomenologically. For the other two operators, on the other hand, the two solutions $P_A$ and $P_C$ result in the same (tiny) values of coefficients but with opposite signs.

\section{Effects of solutions $P_A$ and $P_C$ in $B_c^-\to\gamma\tau^-\bar\nu_\tau$, $\bar B\to D^{(\ast)}\tau\bar\nu_\tau$ and $B\to X_c\tau\bar\nu_\tau$ decays}
\label{sec:numerical analysis}

In this section, we study the effects of solutions $P_A$ and $P_C$ in $B_c^-\to\gamma\tau^-\bar\nu_\tau$, $\bar B\to D^{(\ast)}\tau\bar\nu_\tau$ and $B\to X_c\tau\bar\nu_\tau$ decays. All the relevant analytic formulae for these decays are collected in Appendixes~A and B. In Table~\ref{tab:inputs}, we list the input parameters used in our numerical analyses, with all the other ones taken from the Particle Data Group~\cite{Agashe:2014kda}.

\begin{table}[t]
\setlength{\abovecaptionskip}{0pt}
\setlength{\belowcaptionskip}{10pt}
\begin{center}
\caption{\label{tab:inputs} Input parameters used in our numerical analyses.}
\renewcommand\arraystretch{1.5}
\tabcolsep 0.25in
\begin{tabular}{ccc}
\toprule
\toprule
Parameter & Value & Reference(s)\\
\midrule
$\alpha_s(M_Z)$ & $0.1185\pm0.0006$ & \cite{Agashe:2014kda}\\
$\alpha_e$ & $1/128$ & \cite{Agashe:2014kda}\\
$m_t$ & $(173.21\pm0.87)~{\rm GeV}$ & \cite{Agashe:2014kda}\\
$m_b(m_b)$ & $(4.18\pm0.03)~{\rm GeV}$ & \cite{Agashe:2014kda}\\
$m_c(m_c)$ & $(1.275\pm0.025)~{\rm GeV}$ & \cite{Agashe:2014kda}\\
\addlinespace
$\tau_{B^+}$        & $(1.638\pm 0.004)~{\rm ps}$ & \cite{Agashe:2014kda}\\
$\tau_{B^0}$        & $(1.520\pm 0.004)~{\rm ps}$ & \cite{Agashe:2014kda}\\
$|V_{cb}|$        & $(41.1\pm 1.3)\times10^{-3}$ & \cite{Agashe:2014kda}\\
\addlinespace
$\rho_D^2$ & $1.086\pm0.070$ & \cite{Aubert:2008yv,Aubert:2009ac,Glattauer:2014}\\
$V_1(1)$ & $0.908\pm0.017$ & \cite{Okamoto:2004xg}\\
\addlinespace
$\rho_{D^\ast}^2$ & $1.207\pm0.026$   & \cite{Amhis:2014hma}\\
$h_{A_1}(1)$    & $0.908\pm0.017$   & \cite{Bailey:2010gb}\\
$R_1(1)$ & $1.406\pm0.033$ & \cite{Amhis:2014hma}\\
$R_2(1)$ & $0.853\pm0.020$ & \cite{Amhis:2014hma}\\
\bottomrule
\bottomrule
\end{tabular}
\end{center}
\end{table}

We give our predictions for the observables in $B_c^-\to\gamma\tau^-\bar\nu_\tau$, $\bar B\to D^{(\ast)}\tau\bar\nu_\tau$ and $B\to X_c\tau\bar\nu_\tau$ decays both within the SM as well as in the one scalar leptoquark scenario with the operator parameters taking the values labelled by $P_A$ and $P_C$ in Eq.~\eqref{eq:parameter}. The theoretical uncertainty for an observable is evaluated by varying each input parameter within its corresponding allowed ranges and then adding the individual uncertainties in quadrature~\cite{Hocker:2001xe,Charles:2004jd,Li:2013vlx}.

\subsection{$B_c^-\to \gamma\tau^-\bar\nu_\tau$}

Firstly, we present our predictions for the branching ratio by just giving their center values, which are shown in Table~\ref{tab:Bc radiative}. In our calculation, we adopt the cut condition for the photon energy $E_\gamma\geq1{\rm GeV}$, and take the constituent quark mass values $m_b=4.8~{\rm GeV}$, $m_c=1.5~{\rm GeV}$~\cite{Qiao:2012hp,Wang:2015bka}.

\begin{table}[t]
  \setlength{\abovecaptionskip}{0pt}
  \setlength{\belowcaptionskip}{10pt}
  \begin{center}
  \caption{\label{tab:Bc radiative}Branching ratios of $B_c^-\to\gamma\tau^-\bar\nu_\tau$ both within the SM and in the $P_A$ and $P_C$ cases.}
  \renewcommand\arraystretch{1.5}
  \tabcolsep 0.25in
  \begin{tabular}{cccc}
  \toprule
  \toprule
  Observable & SM & $P_A$ & $P_C$ \\
  \midrule
  $\mathcal B(B_c^-\to\gamma\tau^-\bar\nu_\tau)\times10^5$ & 2.36 & 3.01 & 2.94  \\
  \bottomrule
  \bottomrule
  \end{tabular}
  \end{center}
\end{table}

From Table~\ref{tab:Bc radiative}, one can see that the branching ratios are enhanced by about $27\%$ in both the $P_A$ and $P_C$ cases, relative to our SM prediction, which is in agreement with the ones in the literatures~\cite{Chang:1997re,Chiladze:1998ny,Colangelo:1999gb,Lih:1999it,Chang:1999gn,
Barik:2008zz,Wang:2015bka}.

\begin{figure}[t]
  \centering
  \includegraphics[width=0.48\textwidth]{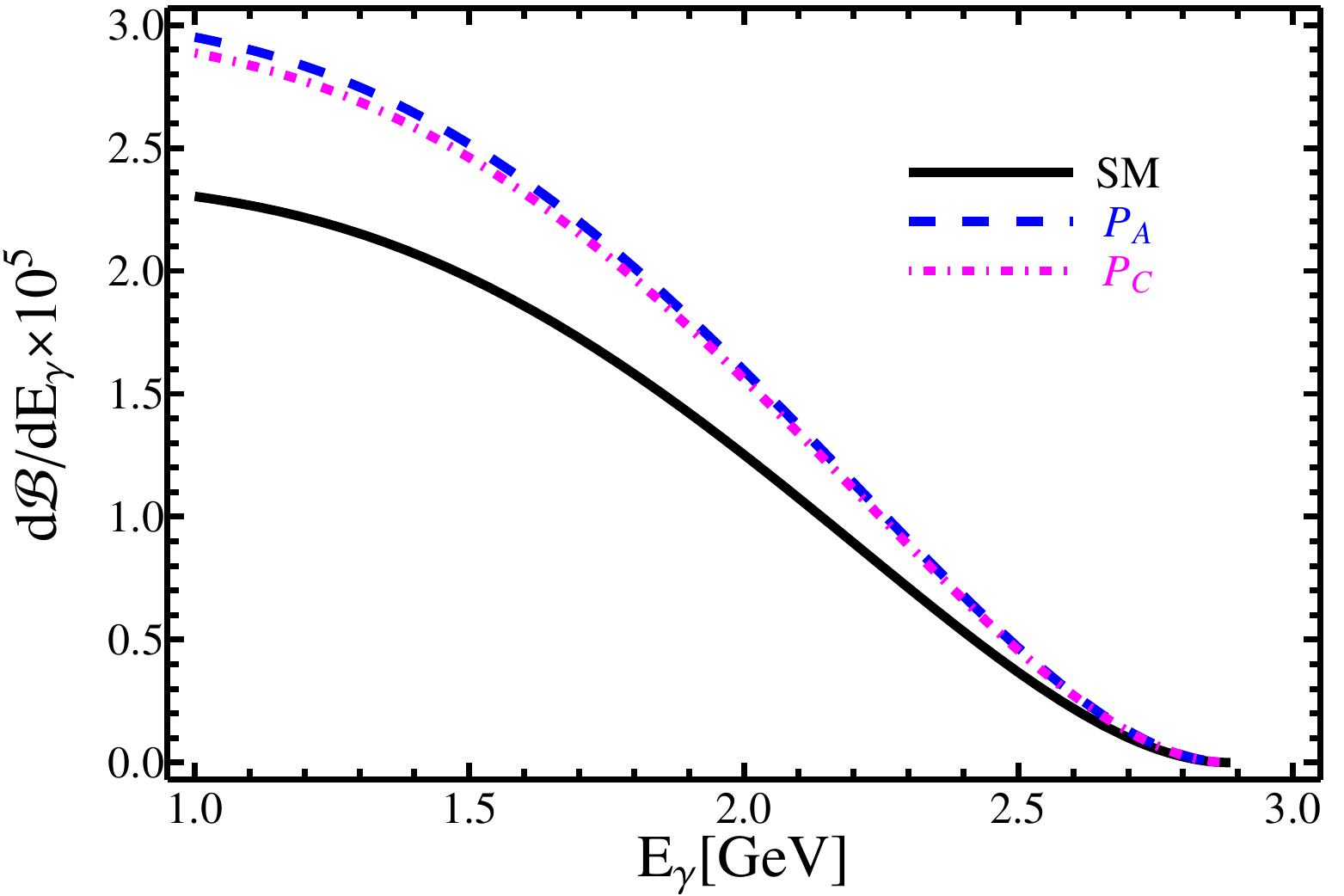}
  \caption{\label{fig:Bc spectrum} The dependence of the differential branching ratios on the photon energy $E_\gamma$.}
\end{figure}

In Fig.~\ref{fig:Bc spectrum}, we present the dependence of the differential branching ratios on the photon energy $E_\gamma$. One can see that the effects of solutions $P_A$ and $P_C$ are both significant in the region $1~{\rm GeV}<E_\gamma<1.5~{\rm GeV}$, but become tiny near the end point $E_\gamma=(m_{B_c}^2-m_\tau^2)/(2m_{B_c})$. While being enhanced both in the $P_A$ and $P_C$ cases, the predicted differential branching ratios coincide almost with each other and are therefore indistinguishable.

It is well-known that, while the width for a purely leptonic decay of a charged pseudoscalar meson is helicity suppressed by $m_{\ell}^2/m_{P^+}^2$~($m_{P^+}$ is the meson mass), the corresponding radiative leptonic decay relieves the helicity suppression, and enhances the decay rate, especially for $\ell=e,\mu$, at expense of much larger theoretical uncertainties~\cite{Braun:2012kp,Beneke:2011nf,DescotesGenon:2002mw,Korchemsky:1999qb}. For the $B_c^-\to (\gamma) \tau^-\bar\nu_\tau$ decays, however, this is not the case; since $B_c^-\to \tau^-\bar\nu_\tau$ does not suffer so much from the helicity suppression, the photons radiated from heavy quarks and heavy $\tau$ do not enhance the decay rate, and the resulting extra electromagnetic coupling $\alpha_e$ will suppress $B_c^-\to\gamma\tau^-\bar\nu_\tau$~\cite{Chang:1997re,Chiladze:1998ny,Colangelo:1999gb,Lih:1999it,Chang:1999gn,
Barik:2008zz,Wang:2015bka}.

Together with the lattice QCD calculation of the decay constant $f_{B_c}$~\cite{Colquhoun:2015oha}, $\mathcal B(B_c^-\to \tau^-\bar\nu_\tau)$ could be reliably predicted. To test the one scalar leptoquark scenario, the purely leptonic decay $B_c^-\to\tau^-\bar\nu_\tau$ is very powerful, especially if LHCb could measure the branching ratio $\mathcal B(B_c^-\to\tau^-\bar\nu_\tau)$ with a precision of $5\%$, since the remaining two best-fit solutions $P_A$ and $P_C$ just enhance it by $10\%$ and $8\%$, respectively, as shown in Eq.~\eqref{eq:Bc result}. However, unlike the measurements of $B_u^-\to\tau^-\bar\nu_\tau$ at BaBar and Belle operated at the $\Upsilon(4S)$ resonance with $B_u^{\pm}$ produced in pairs, it would be extremely difficult to measure $B_c^-\to\tau^-\bar\nu_\tau$ at LHCb, because the presence of at least two neutrinos per $B_c^-$ decay and the inability to impose kinematic constraints on the center-of-mass energy make background rejection incredibly challenging. The radiative mode would face even more challenges from vetoing photons from excited $B_c^{\ast}\to B_c \gamma$ decays. While being very challenging, it is worthwhile for LHCb to make delicately  experimental studies of these decays thanks to the high luminosity and the large $B_c^-$ production cross-section at the LHC~\cite{Gouz:2002kk}.

\subsection{$\bar B\to D^{(\ast)}\tau\bar\nu_\tau$}

In this subsection, we present firstly in Table~\ref{tab:RD} our predictions for the ratios $R(D^{(\ast)})$ and the branching fractions $\mathcal B(\bar B\to D^{(\ast)}\tau\nu_\tau)$, both within the SM and in the $P_A$ and $P_C$ cases. One can see from the table that the values of $R(D^{(\ast)})$ in both the $P_A$ and $P_C$ cases coincide very well with the experimental data, as it should be.

\begin{table}[t]
  \setlength{\abovecaptionskip}{0pt}
  \setlength{\belowcaptionskip}{10pt}
  \begin{center}
  \caption{\label{tab:RD} Theoretical and experimental values of $R(D^{(\ast)})$ and $\mathcal B(\bar B\to D^{(\ast)}\tau\nu_\tau)$~(in unit of $10^{-2}$), both within the SM and in the $P_A$ and $P_C$ cases. The statistical and systematic uncertainties have been added in quadrature for the experimental data.}
  \renewcommand\arraystretch{1.5}
  \tabcolsep 0.12in
  \begin{tabular}{ccccc}
  \toprule
  \toprule
  Observable   & SM                    & $P_A$   & $P_C$  & Exp\\
  \midrule
  $R(D)$       & $0.298\pm0.009$   & $0.388\pm0.012$   & $0.380\pm0.011$ &  $0.397\pm0.049$~\cite{HFAG:2016}\\
  $R(D^\ast)$  & $0.253\pm0.002$             & $0.325\pm0.002$   & $0.318\pm0.002$ & $0.316\pm0.019$~\cite{HFAG:2016}\\ \addlinespace
  $\mathcal B(\bar B\to D\tau\nu)$ & $0.72\pm0.06$ & $0.94 \pm0.07$ & $0.92\pm0.07$ & $1.07\pm0.18$~\cite{Agashe:2014kda}\\
  $\mathcal B(\bar B\to D^\ast\tau\nu)$ & $1.30\pm0.04$& $1.67\pm0.05$& $1.64\pm0.04$ & $1.64\pm0.15$~\cite{Agashe:2014kda}\\
  \bottomrule
  \bottomrule
  \end{tabular}
  \end{center}
\end{table}

\begin{figure}[t]
  \centering
  \subfigure[]{\includegraphics[width=3.0in]{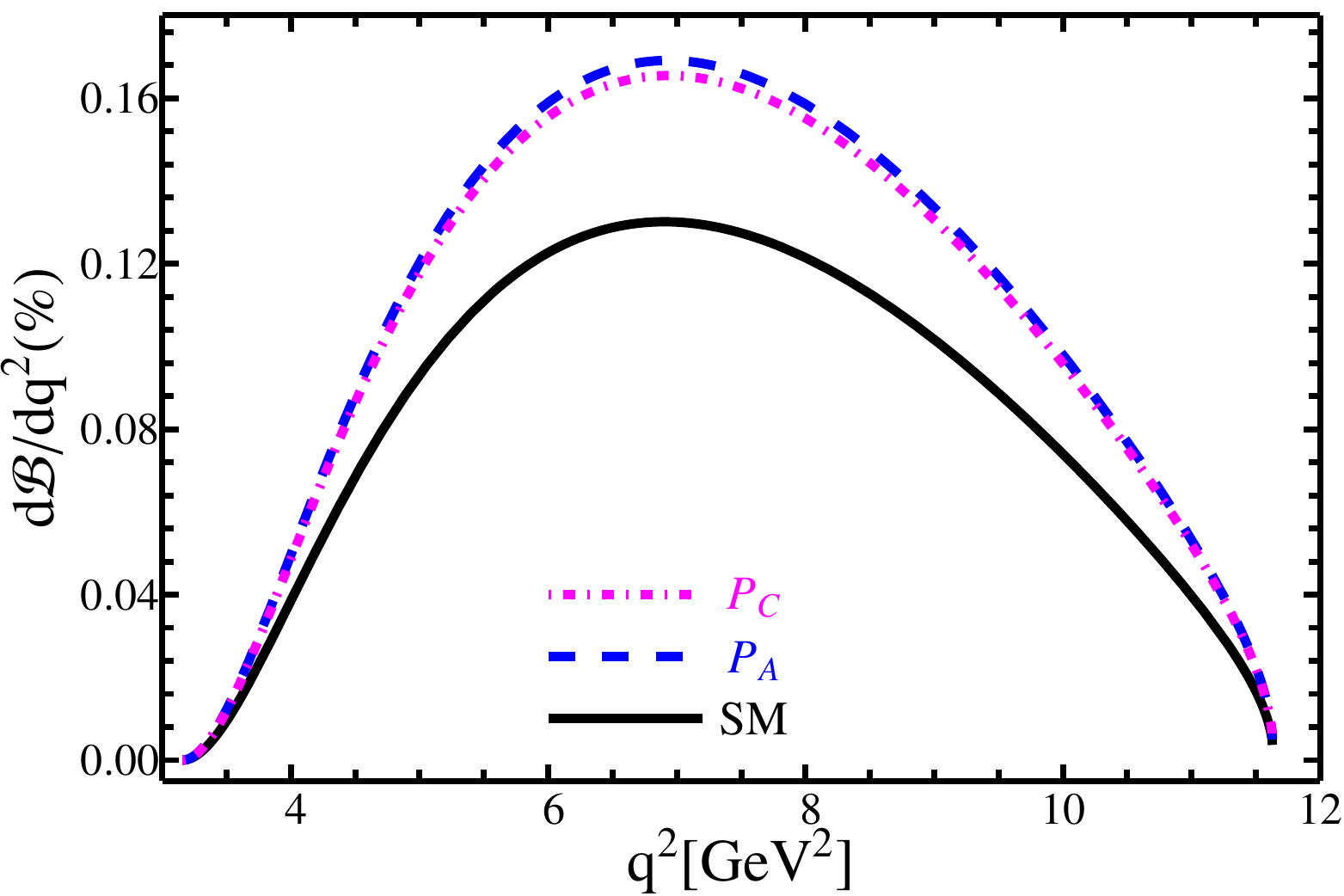}}
  \hspace{0.2in}
  \subfigure[]{\includegraphics[width=3.0in]{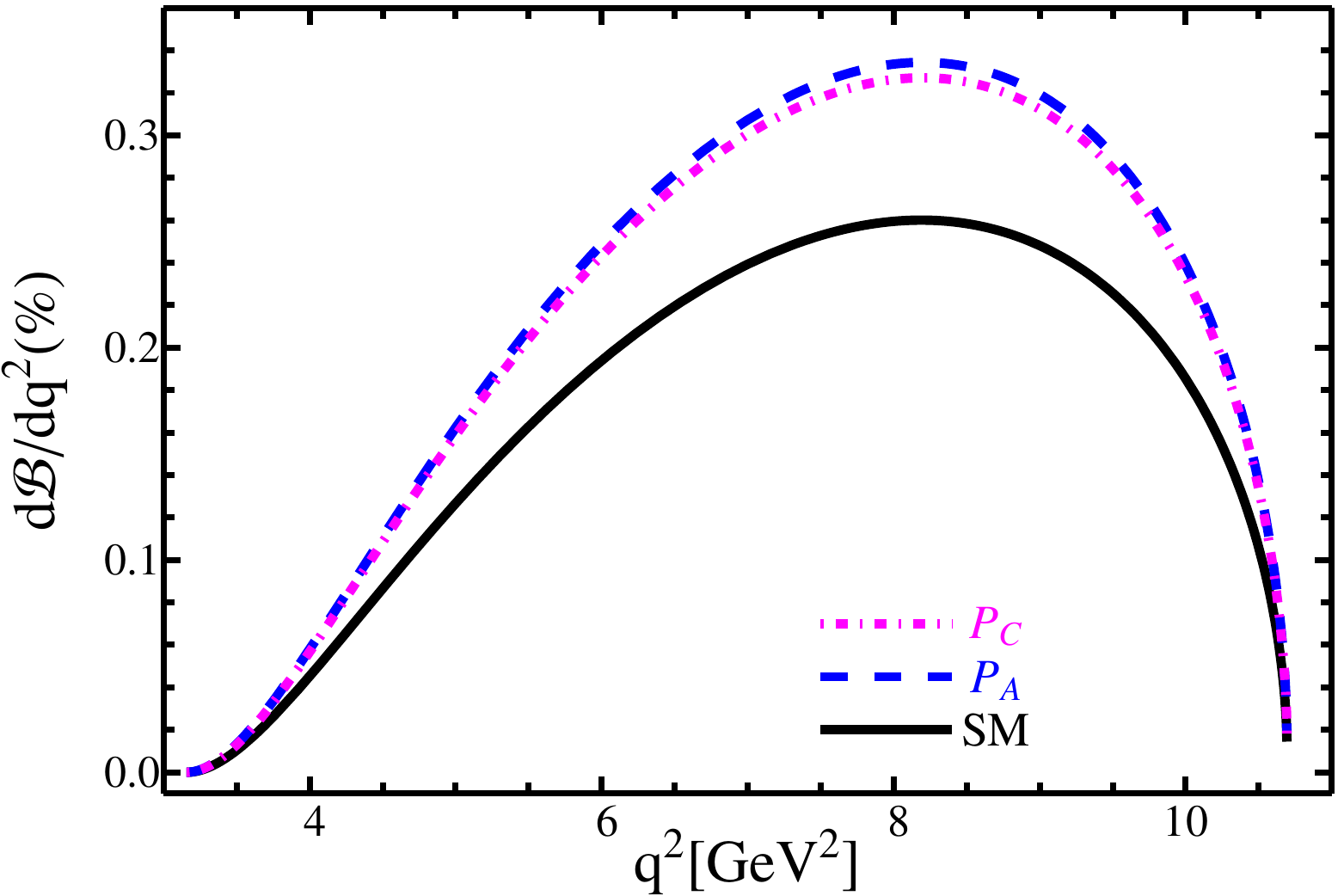}}
   \caption{\label{fig:dbr} The $q^2$ distributions of the differential branching fractions for $\bar B\to D\tau\bar{\nu}_{\tau}$~(a) and for $\bar B\to D^*\tau\bar{\nu}_{\tau}$~(b) decays.}
\end{figure}

\begin{figure}[t]
  \centering
  \subfigure[]{\includegraphics[width=3.0in]{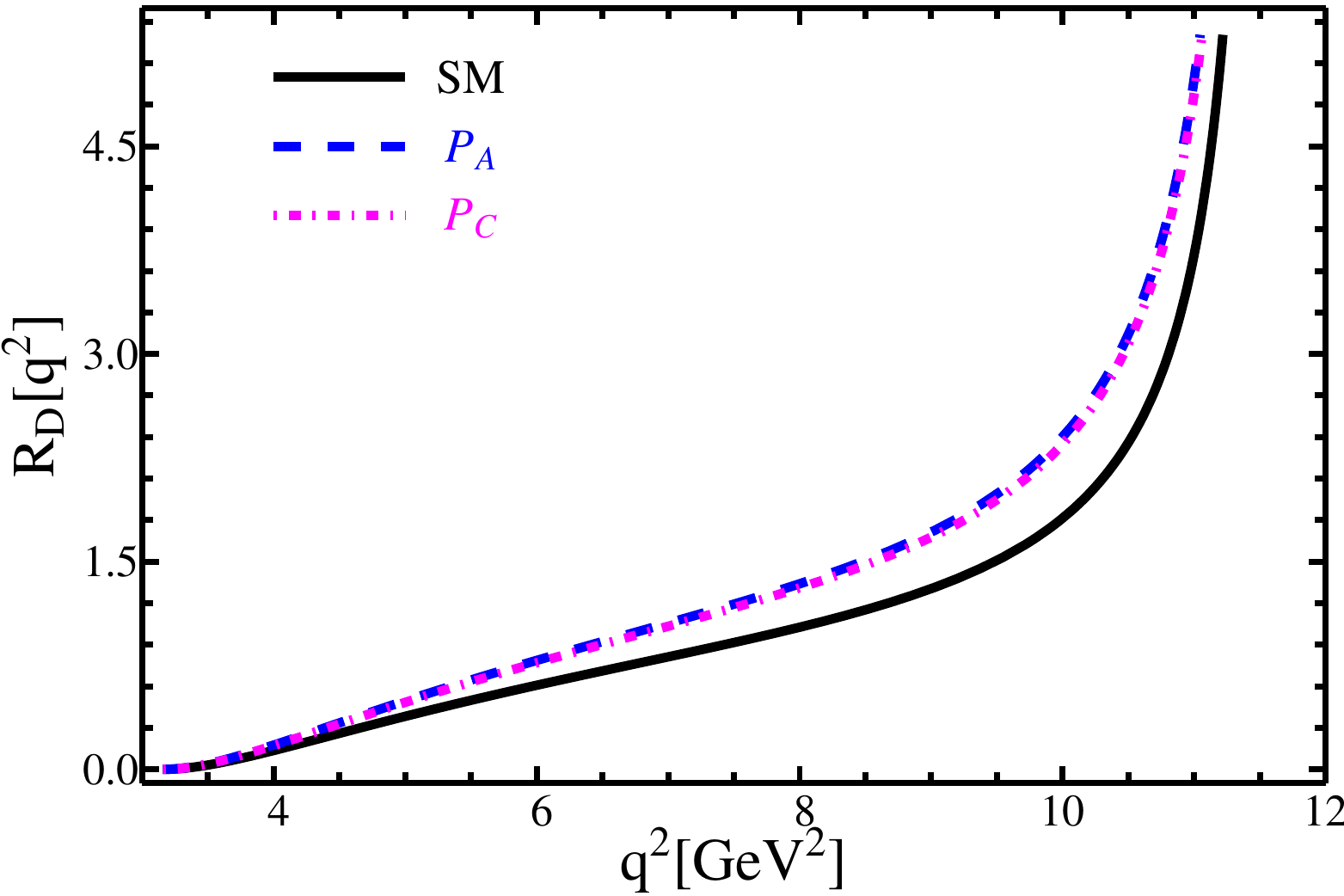}}
  \hspace{0.2in}
  \subfigure[]{\includegraphics[width=3.0in]{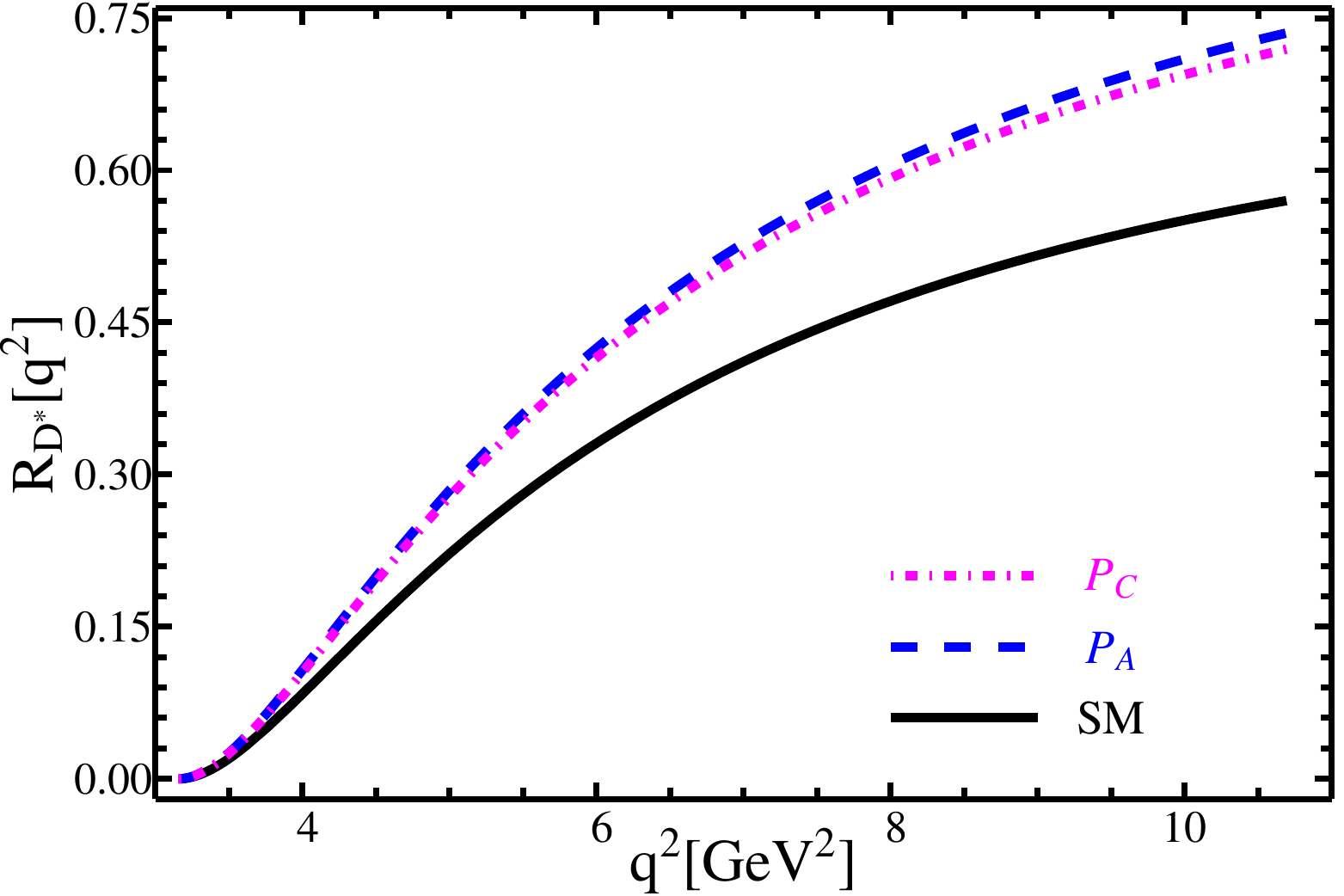}}
  \caption{\label{fig:RD} The $q^2$ distributions of the ratios $R_D(q^2)$~(a) and $R_{D^\ast}(q^2)$~(b).}
\end{figure}

\begin{figure}[hb]
  \centering
  \subfigure[]{\includegraphics[width=3.0in]{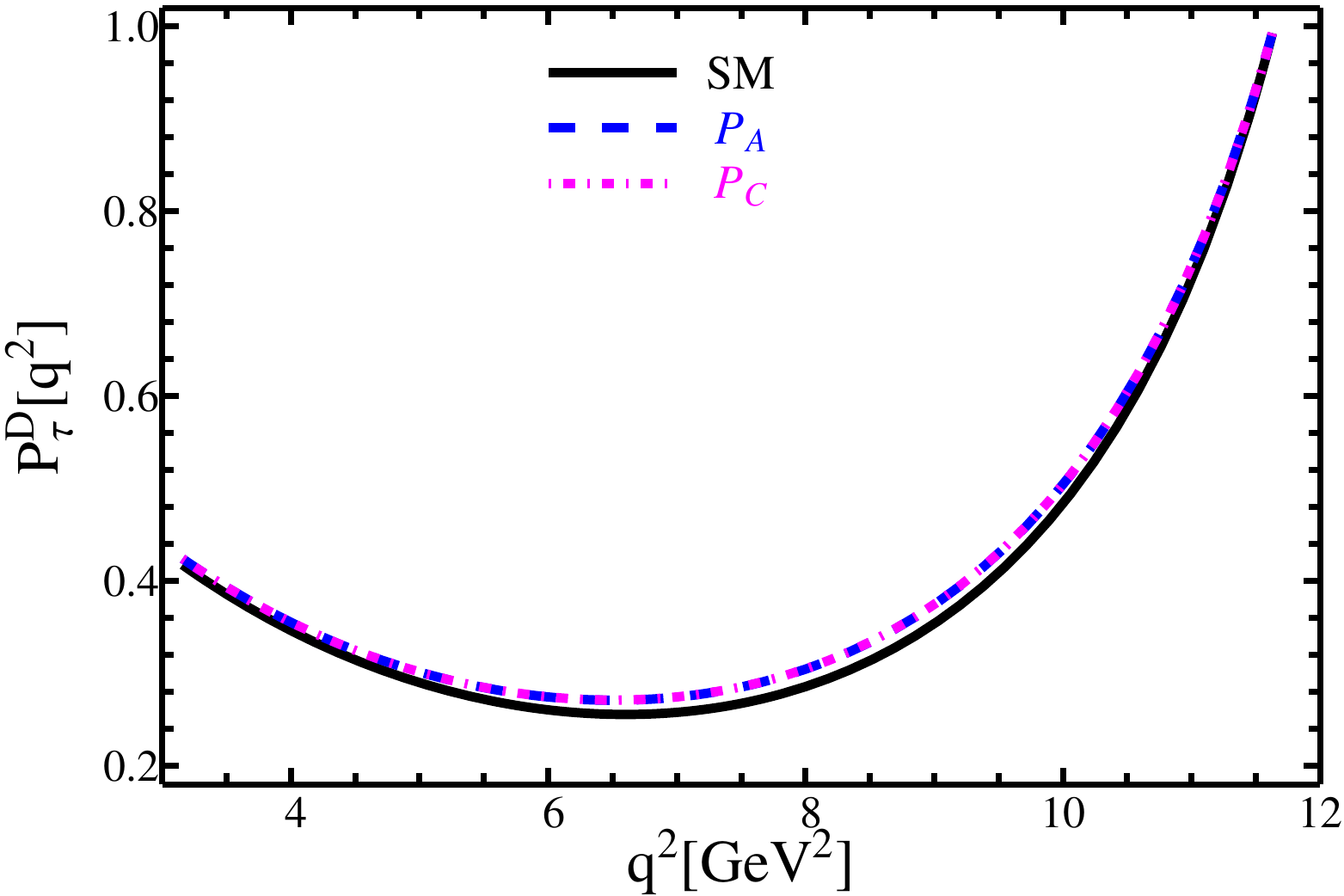}}
  \hspace{0.2in}
  \subfigure[]{\includegraphics[width=3.0in]{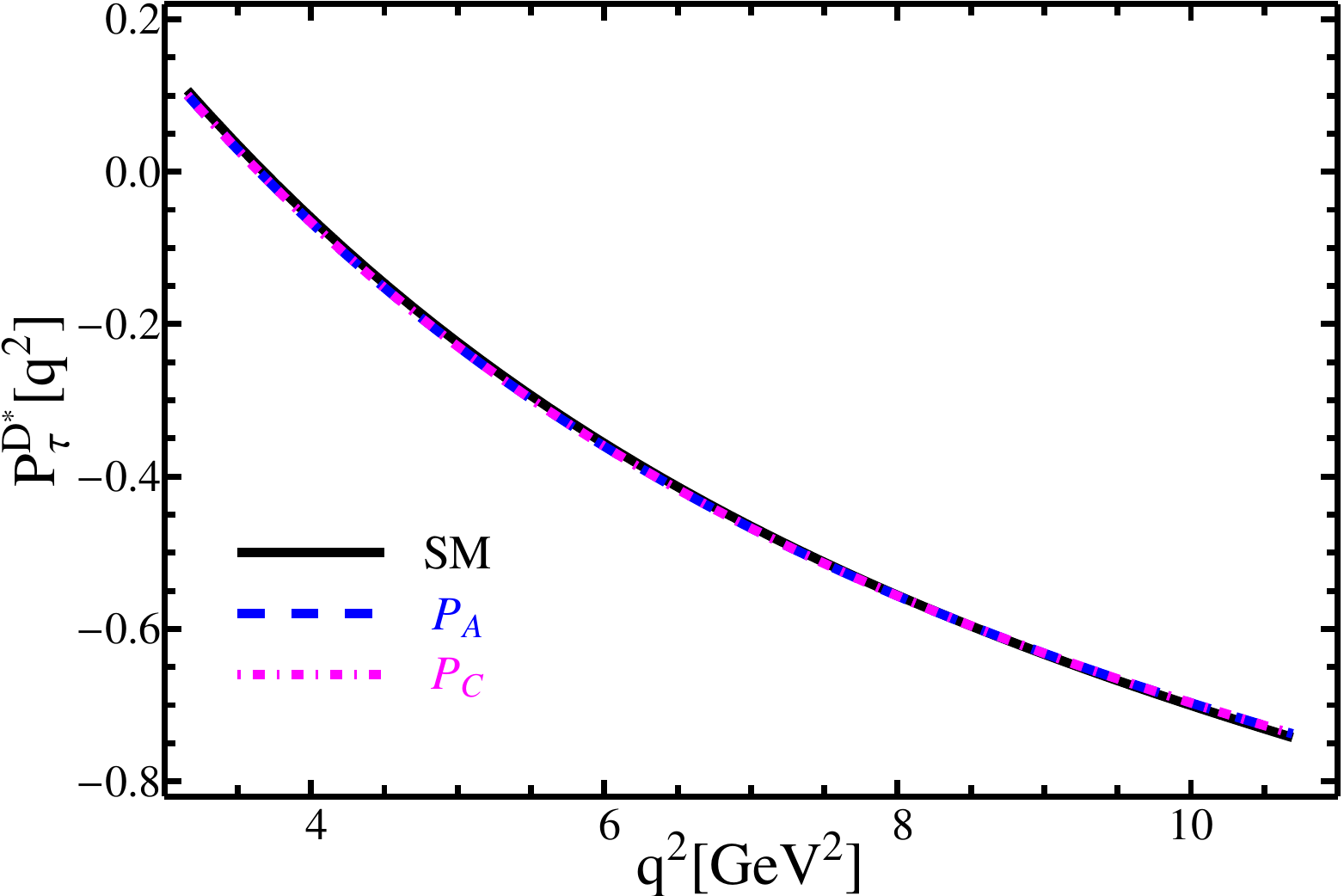}}
  \caption{\label{fig:ptau} The $q^2$ distributions of the $\tau$ longitudinal  polarization in $\bar B\to D\tau\bar{\nu}_{\tau}$~(a) and in $\bar B\to D^\ast\tau\bar{\nu}_{\tau}$~(b) decays.}
\end{figure}

\begin{figure}[t]
  \centering
  \subfigure[]{\includegraphics[width=2.15in]{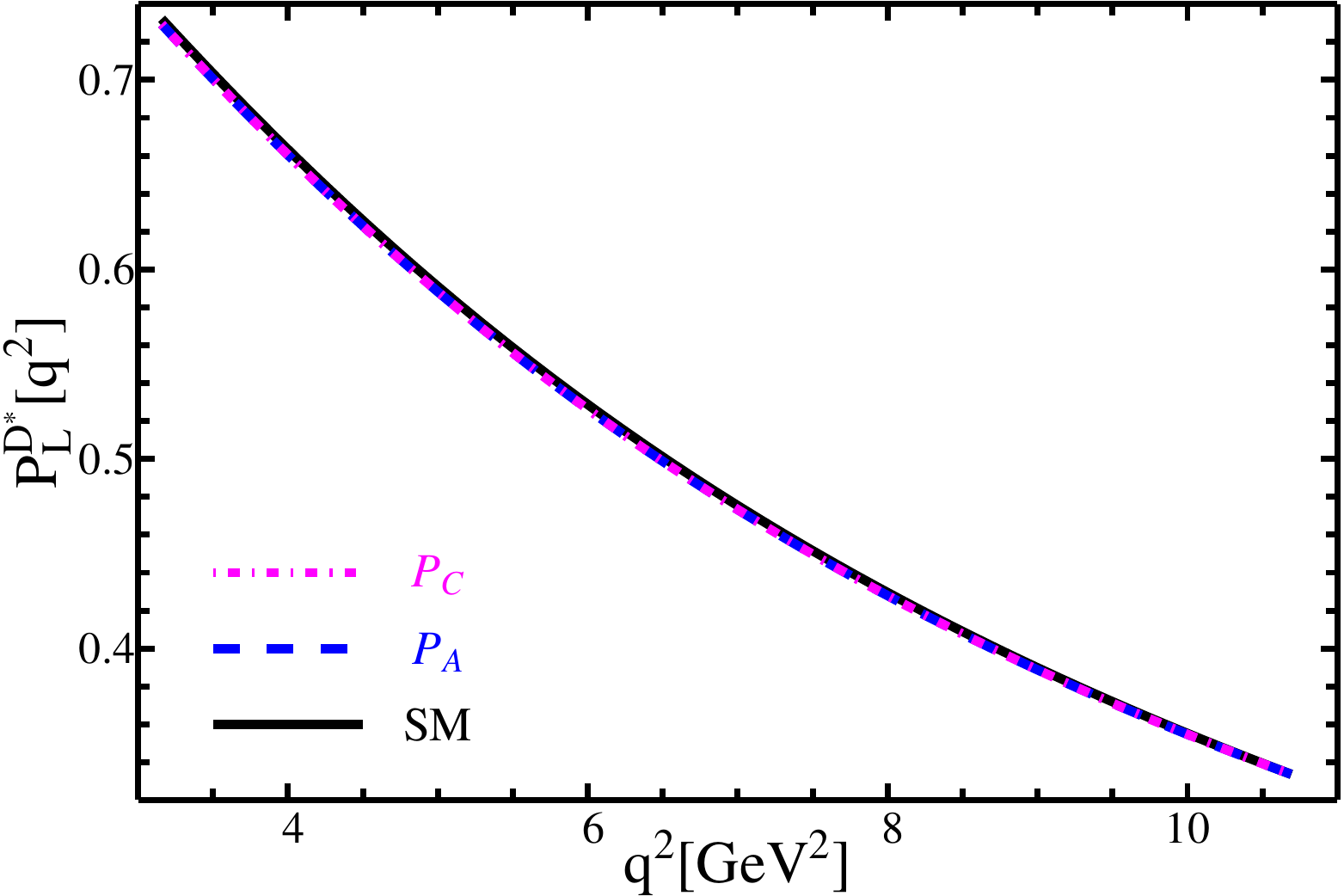}}
  \hspace{0.01in}
  \subfigure[]{\includegraphics[width=2.15in]{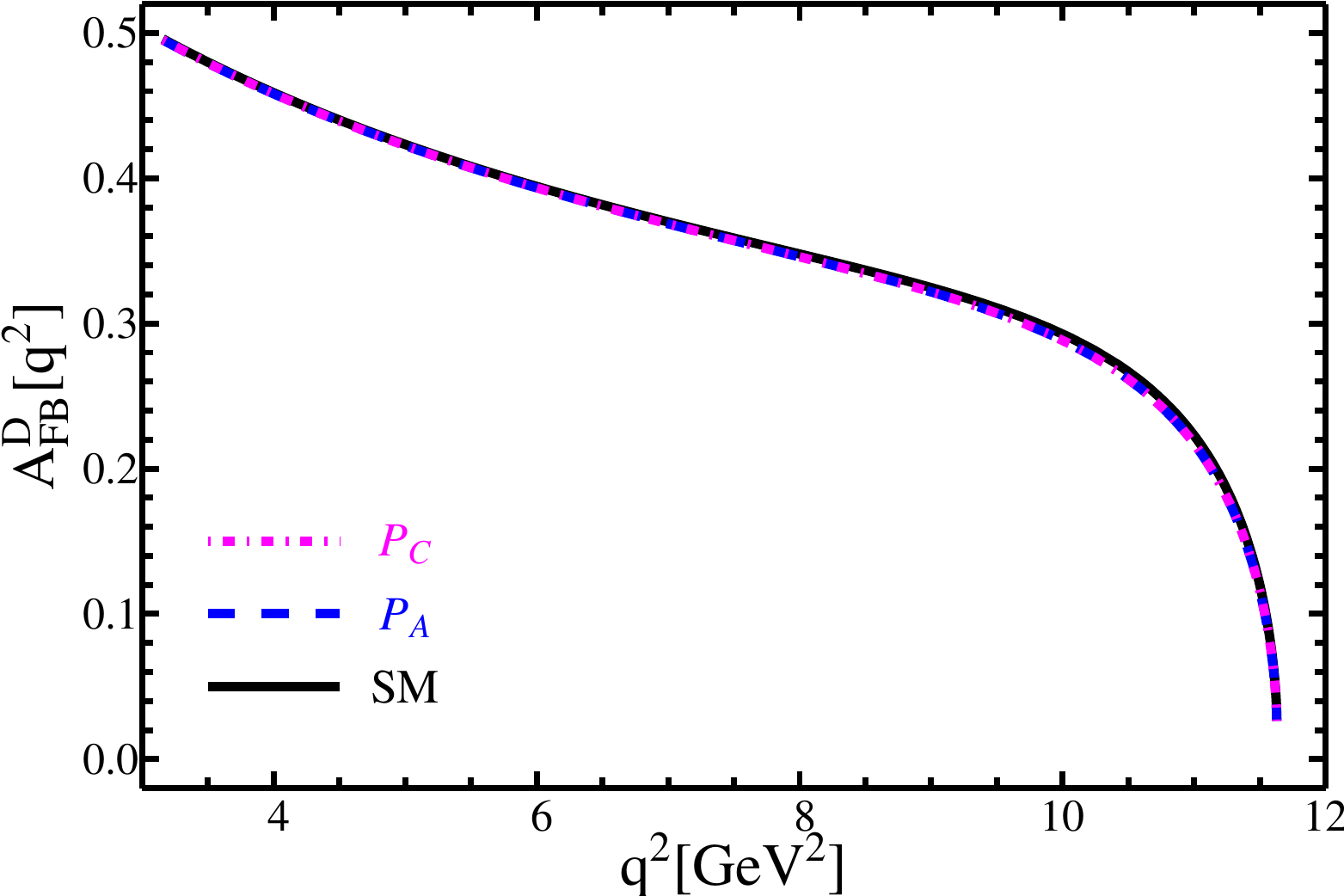}}
  \hspace{0.01in}
  \subfigure[]{\includegraphics[width=2.15in]{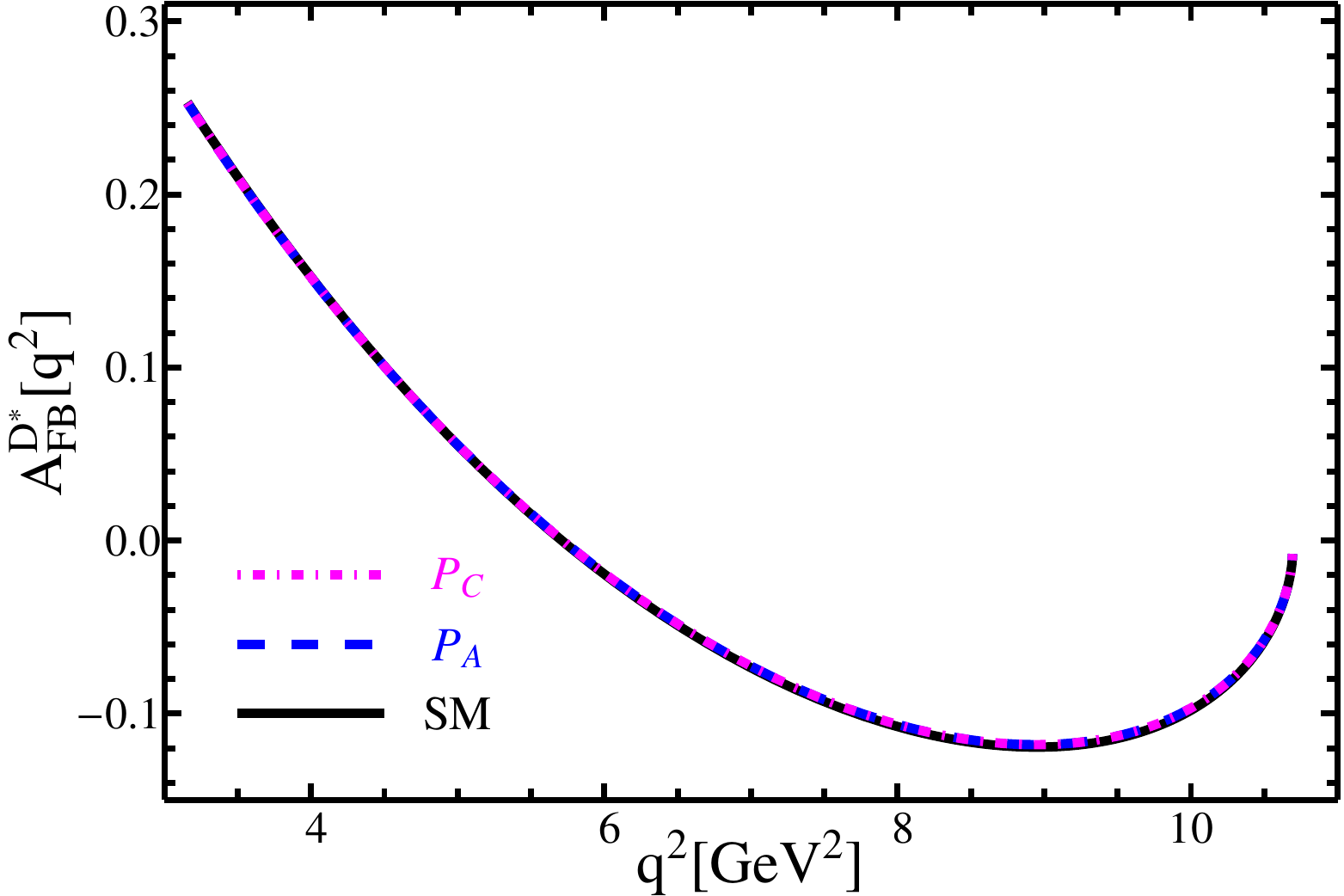}}
  \caption{\label{fig:PDst and AFB} The $q^2$ distributions of the $D^\ast$ longitudinal polarization in $\bar B\to D^\ast\tau\bar{\nu}_{\tau}$~(a), and that of the lepton forward-backward asymmetries in $\bar B\to D\tau\bar{\nu}_{\tau}$~(b) and in $\bar B\to D^\ast\tau\bar{\nu}_{\tau}$~(c) decays.}
\end{figure}

We now analyze in turn the $q^2$ distributions of the differential branching fractions~(shown in Fig.~\ref{fig:dbr}), the ratios $R_{D^{(\ast)}}(q^2)$~(Fig.~\ref{fig:RD}), the polarizations of $\tau$~(Fig.~\ref{fig:ptau}) and $D^\ast$~(Fig.~\ref{fig:PDst and AFB}~(a)), as well as the lepton forward-backward asymmetry defined as the relative difference between the partial decay rates where the angle $\theta$ between $D^{(\ast)}$ and $\tau$ three-momenta in the $\tau$-$\bar\nu_\tau$ center-of-mass frame is greater or smaller than $\pi/2$~(cf. Eq.~\eqref{eq:AFB})~(Figs.~\ref{fig:PDst and AFB}~(b) and \ref{fig:PDst and AFB}~(c)) in $\bar B\to D^{(\ast)}\tau\bar\nu_\tau$ decays. For simplicity, we plot only the central values of these observables at each $q^2$ point. From these figures, we make the following observations:
\begin{itemize}
\item As shown in Fig.~\ref{fig:dbr}, the differential branching ratio ${\rm d}\mathcal B(\bar B\to D\tau\bar{\nu}_{\tau})/{\rm d}q^2$ is largely enhanced around $q^2\sim 7~{\rm GeV}^2$, while ${\rm d}\mathcal B(\bar B\to D^\ast\tau\bar{\nu}_{\tau})/{\rm d}q^2$ around $8~{\rm GeV}^2$. Furthermore, both $P_A$ and $P_C$ predict similar $q^2$ behaviors as in the SM for these two observables. As the measured differential distributions by BaBar~\cite{Lees:2013uzd} and Belle~\cite{Huschle:2015rga,Abdesselam:2016cgx} are still quite uncertain, it is currently unable to discriminate the NP from the SM predictions. More precise measurements of these observables by LHCb and Belle-II are, therefore, very necessary.

\item From Fig.~\ref{fig:RD}, one can see that the scalar leptoquark effects provide overall enhancements for both $R_D(q^2)$ and $R_{D^\ast}(q^2)$ in the whole kinematic region. However, the enhancement is small for $R_D(q^2)$, but quite large for $R_{D^\ast}(q^2)$ in the large $q^2$ region. This could be tested at Belle-II in the near future.

\item As shown in Figs.~\ref{fig:ptau} and \ref{fig:PDst and AFB}, for the $\tau$ and $D^{\ast}$ longitudinal polarizations, as well as the lepton forward-backward asymmetries in these decays, results obtained in the $P_A$ and $P_C$ cases coincide not only with each other, but also with the corresponding SM predictions. This is naively what should be expected, because the scalar leptoquark effects appear both in the numerator and in the denominator of these observables and are cancelled to a large extent, making these observables almost independent of their contributions.
\end{itemize}

\subsection{$B\to X_c\tau^-\bar\nu_\tau$}

Finally, we consider the inclusive semi-leptonic $B$-meson decays. Similar to the case in exclusive decays, we can also define a ratio $R(X_c)$ for inclusive decay rates,
\begin{equation}
R(X_c)=\frac{\mathcal B(\bar B\to X_c\tau\bar\nu_\tau)}{\mathcal B(\bar B\to X_c\ell\bar\nu_\ell)}\,,\qquad \ell=e/\mu\,,
\end{equation}
which can be calculated precisely with an operator product expansion~\cite{Manohar:2000dt,Ligeti:2014kia,Falk:1994gw,Chay:1990da,Shifman:1984wx}. With the most recent world average $\mathcal B(B^-\to X_ce\bar\nu_e)=(10.92\pm0.16)\%$~\cite{Bernlochner:2012bc,Amhis:2014hma}, one can then get the prediction for $\mathcal B(B^-\to X_c\tau\bar\nu_\tau)$, free of the large uncertainty due to the factor $m_b^5$. Here we consider neither the $\mathcal O(\alpha_s)$ QCD nor the $\mathcal O(\Lambda_{\rm QCD}/m_b)$ power corrections, and take the heavy quark on-shell masses with $m_b=4.6~{\rm GeV}$ and $m_c=1.15~{\rm GeV}$~\cite{Gambino:2015ima}.

Our numerical results of the ratio $R(X_c)$ and the branching fraction $\mathcal B(B^-\to X_c\tau\bar\nu_\tau)$ are given in Table~\ref{tab:inclusive}. From the table, one can see that both $R({X_c})$ and $\mathcal B(B^-\to X_c\tau\bar\nu_\tau)$ are enhanced by the scalar leptoquark, and our SM value of $R(X_c)$ is roughly consistent with the recent update within the 1S short-distance mass scheme~\cite{Ligeti:2014kia,Freytsis:2015qca}, $R(X_c)=0.223\pm0.004$, obtained with both the $\mathcal O(\Lambda_{\rm QCD}^2/m_b^2)$ and the two-loop QCD corrections included~\cite{Biswas:2009rb}.

\begin{table}[t]
\setlength{\abovecaptionskip}{0pt}
\setlength{\belowcaptionskip}{10pt}
\begin{center}
\caption{\label{tab:inclusive} Predictions for $R(X_c)$ and $\mathcal B(B^-\to X_c\tau\bar\nu_\tau)$ in the SM and in the $P_A$ and $P_C$ cases. }
\renewcommand\arraystretch{1.3}
\tabcolsep 0.12in
\begin{tabular}{cccc}
\toprule
\toprule
Observable       & SM                    & $P_A$      & $P_C$ \\
\midrule
$R(X_c)$         & $0.230$                     & $0.297$   & $0.290$ \\
$\mathcal B(B^-\to X_c\tau^-\bar\nu_\tau)$  & $2.51\%$   & $3.24\%$   & $3.17\%$\\
\bottomrule
\bottomrule
\end{tabular}
\end{center}
\end{table}

\begin{figure}[hb]
  \centering
  \subfigure[]{\includegraphics[width=2.15in]{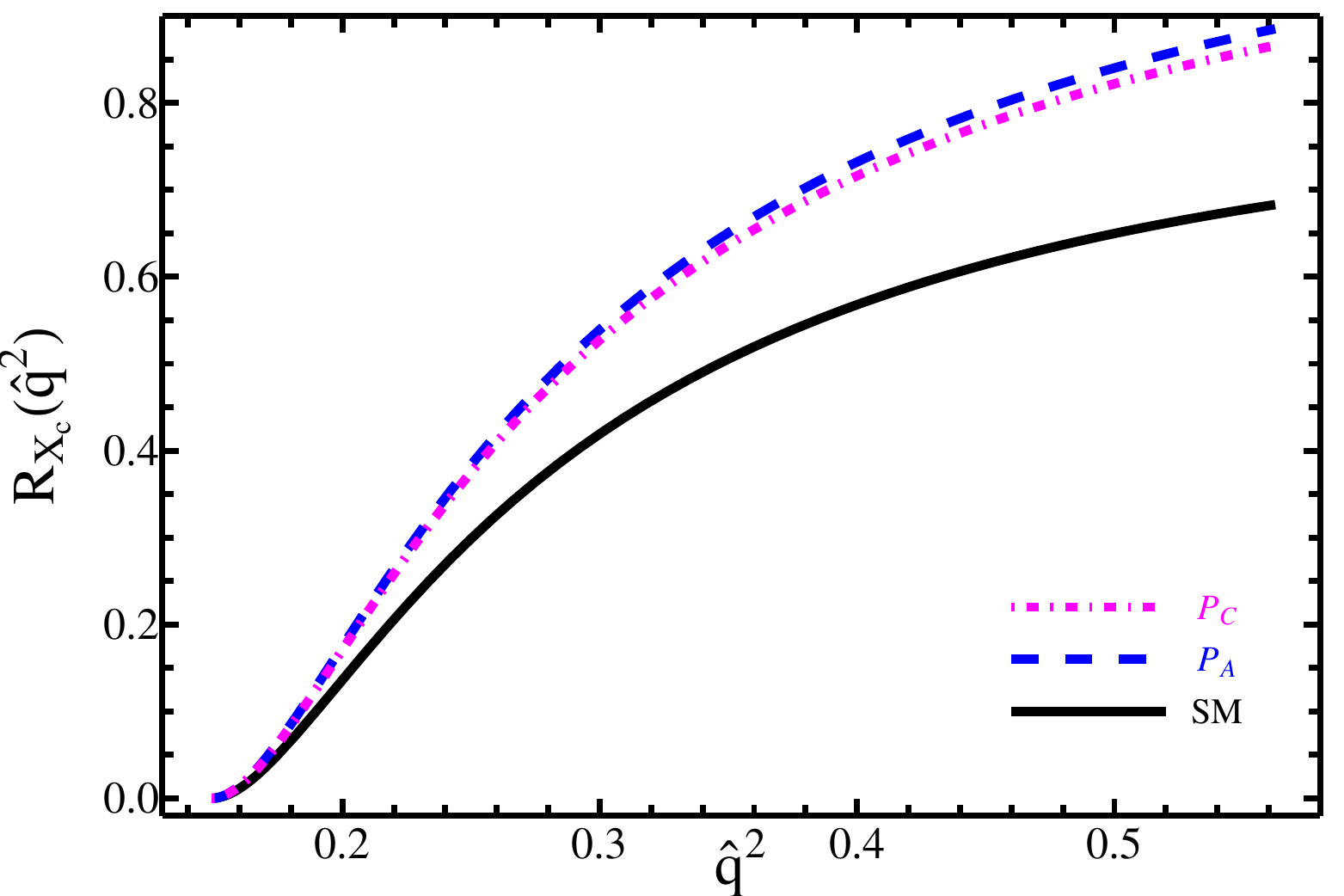}}
  \hspace{0.01in}
  \subfigure[]{\includegraphics[width=2.15in]{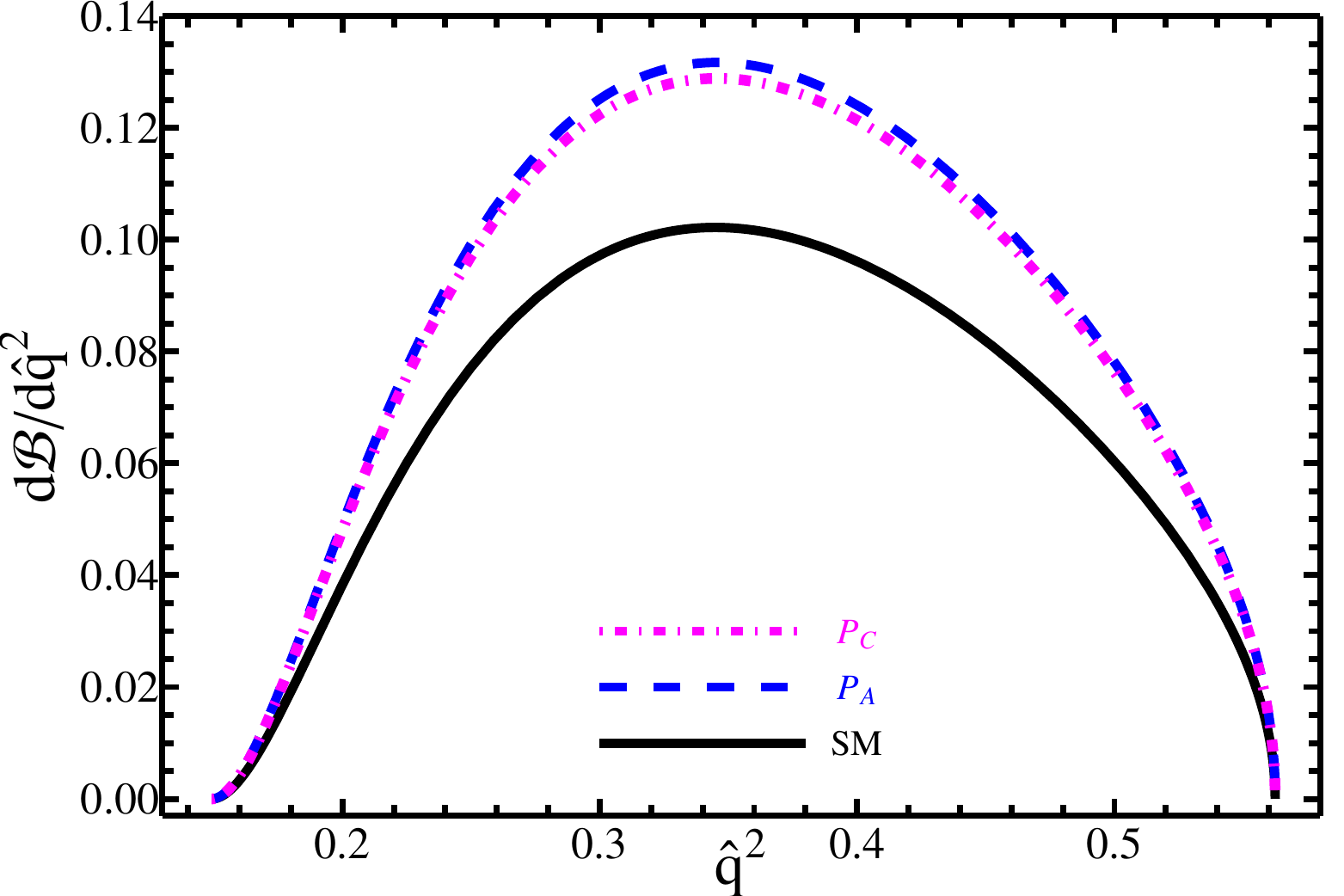}}
  \hspace{0.01in}
  \subfigure[]{\includegraphics[width=2.15in]{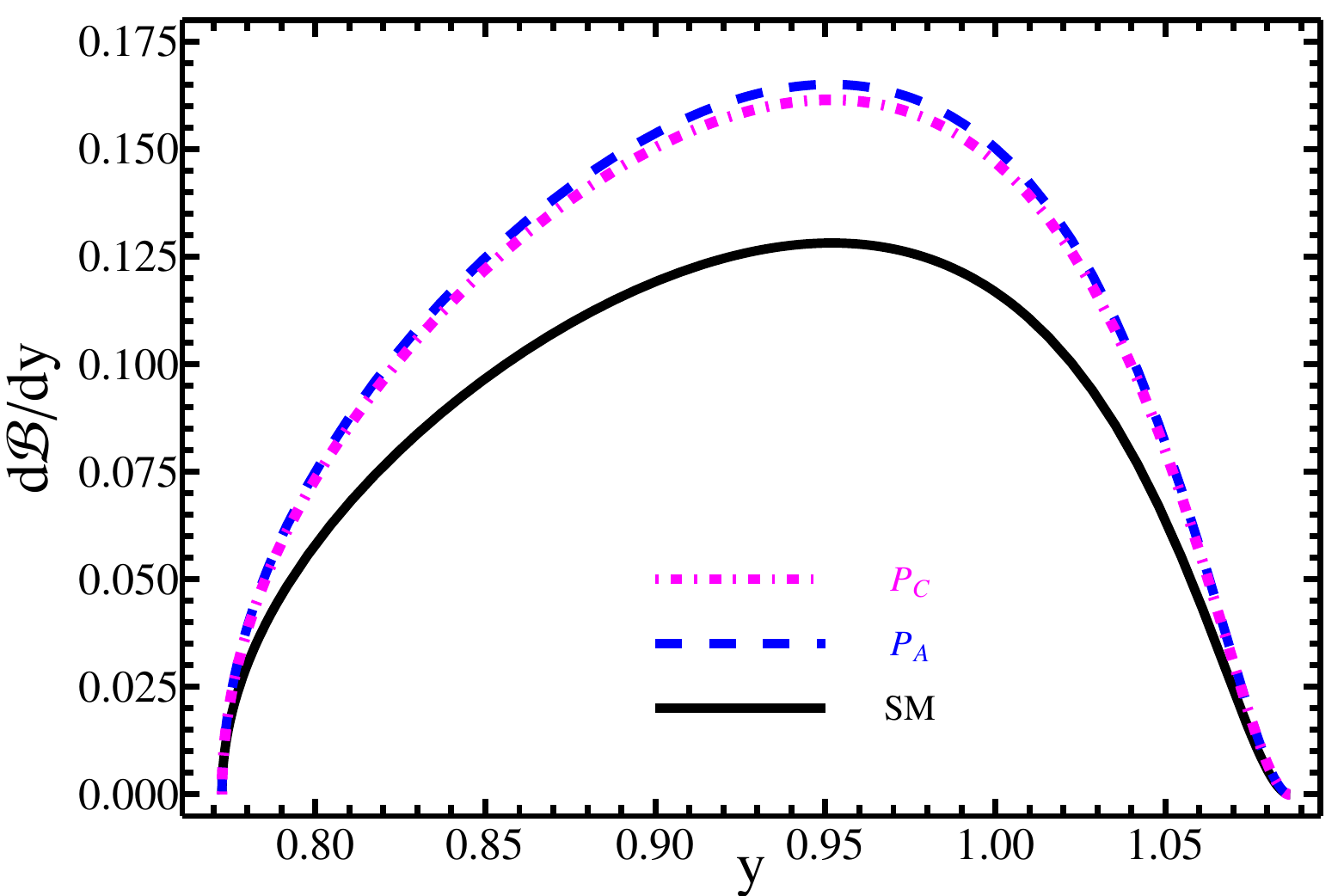}}
  \caption{\label{fig:inclusive} The $\hat q^2$ distributions of the ratio $R_{X_c}(\hat q^2)$~(a) and the differential branching fraction ${\rm d}\mathcal B(B^-\to X_c\tau\bar\nu_\tau)/{\rm d}\hat q^2$~(b), as well as the $\tau$-energy spectrum ${\rm d}\mathcal B(B^-\to X_c\tau\bar\nu_\tau)/{\rm d}y$~(c).}
\end{figure}

In Fig.~\ref{fig:inclusive}, we display the $\hat q^2$ distributions of the ratio $R_{X_c}(\hat q^2)$, the differential branching fraction ${\rm d}\mathcal B(B^-\to X_c\tau\bar\nu_\tau)/{\rm d}\hat q^2$, as well as the differential spectrum of the $\tau$ energy ${\rm d}\mathcal B(B^-\to X_c\tau\bar\nu_\tau)/{\rm d}y$, where $\hat q^2=q^2/m_b^2$ and $y=2E_{\tau}/m_b$\footnote{It should be noted that the lowest-order~(parton-level) prediction for the inclusive spectrum receives substantial corrections from nonperturbative $\Lambda^2_{\rm QCD}/m^2_b$ power corrections and shape-function convolutions in the large $q^2$~(kinematic endpoint) region, which is also the part of the distribution where any reported data will likely be cleanest; for a recently detailed study, see Ref.~\cite{Ligeti:2014kia}.}. One can see that the $\hat q^2$ distributions of these two observables are similar to that of $R_{D^\ast}(q^2)$ and ${\rm d}\mathcal B(\bar B\to D^\ast\tau\bar\nu_\tau)/{\rm d}q^2$, respectively,  since both $R_{X_c}(\hat q^2)$ and ${\rm d}\mathcal B(B^-\to X_c\tau\bar\nu_\tau)/{\rm d}\hat q^2$ are also enhanced by the scalar leptoquark contributions in the whole kinematic region, except for near the origin and the end point regions of $\hat q^2$ for the latter. This is due to the fact that these observables have similar relations with respect to the operator coefficients $C_V$, $C_S$ and $C_T$, which can be seen from Eqs.~\eqref{eq:dBrDst} and \eqref{eq:inclusive width}. It is also found that the $\tau$-energy spectrum shows a different behavior than the differential branching fraction and provides complementary information compared to the latter.

\section{Conclusion}
\label{sec:conclusion}

The anomalies observed in $\bar B\to D^{(\ast)}\tau\bar\nu_\tau$ and $\bar B\to \bar K\ell^+\ell^-$ decays could be resolved with just one scalar leptoquark~\cite{Bauer:2015knc}. Fitting to the current experimental data on the ratios $R(D^{(\ast)})$ and the $q^2$ spectra of $\bar B\to D^{(\ast)}\tau\bar\nu_\tau$, four best-fit solutions denoted by $P_A$, $P_B$, $P_C$ and $P_D$ are obtained~\cite{Freytsis:2015qca}. In this paper, we have explored the possibilities of how to discriminate these four solutions. Firstly, we have shown that two of them, $P_B$ and $P_D$, are already excluded by the purely leptonic decay $B_c^-\to\tau^-\bar\nu_\tau$, because the predicted decay widths by $P_B$ and $P_D$ have already overshot the total width $\Gamma_{B_c}$. The remaining two solutions $P_A$ and $P_C$ would enhance $\mathcal B(B_c^-\to \tau^-\bar\nu_\tau)$ by $10\%$ and $8\%$, respectively. Together with the lattice QCD calculation of the decay constant $f_{B_c}$~\cite{Colquhoun:2015oha}, $\mathcal B(B_c^-\to \tau^-\bar\nu_\tau)$ could be reliably predicted. Given the branching ratio measured to a precision of a few percent at the LHCb, one could then test the interesting one scalar leptoquark model.

By comparing the effects of $P_A$ and $P_C$ at the scale $\mu=m_b$, we find that the two solutions lead to overall different sign of $\mathcal H_{\text{fit}}$, but with just $1.2\%$ difference in the coefficient $C_V^{\text{fit}}$ of $(V-A)\otimes(V-A)$ operator, which is too small to be discriminated from each other phenomenologically. Furthermore, in $\mathcal H_{\text{fit}}$, the coefficient $|C_V^{\text{fit}}|$ is much larger than $|C_S^{\text{fit}}|$ and $|C_T^{\text{fit}}|$, the coefficients of the new scalar and tensor operators, respectively.

Combining these observations and our numerical results, we may draw the following conclusions. The one scalar leptoquark scenario gives nearly the same predictions as in the SM for the $D^\ast$ and $\tau$ longitudinal polarizations and the lepton forward-backward asymmetries in $\bar B\to D^{(\ast)}\tau\bar\nu_\tau$ decays. Although precision measurements of these observables would be very challenging at LHCb and/or Belle-II, any significant deviation from the SM predictions would lead to another model for the $R(D^{(\ast)})$ anomalies. Otherwise, the one scalar leptoquark scenario would be viable and good. For the other observables like $\mathcal B(B_c^-\to \tau^-\bar\nu_\tau)$, $\mathcal B(B_c^-\to \gamma\tau^-\bar\nu_\tau)$, $R_{D^{(\ast)}}(q^2)$, ${\rm d}\mathcal B(\bar B\to D^{(\ast)}\tau\bar\nu_\tau)/{\rm d}q^2$ and $\mathcal B(\bar B\to X_c\tau\bar\nu_\tau)$, on the other hand, the model could generally give sizable enhancements relative to the SM predictions.

With future measurement of $B_c^-\to\tau^-\bar\nu_\tau$ at LHCb and refined measurements of observables in $\bar B\to D^{(\ast)}\tau\bar\nu_\tau$ decays at both LHCb and Belle-II, one could further decipher the various NP models that provide so far good explanations of the $R(D^{(\ast)})$ anomalies.

Finally, we would like to point out that, due to the half-integer-spin of $\Lambda_b$ and $\Lambda_c$ baryons, the semi-leptonic $\Lambda_b \to \Lambda_c \ell \bar\nu_{\ell}$ decays, which are mediated by the same quark-level transition as in $\bar B\to D^{(\ast)}\ell \bar\nu_{\ell}$ decays, can provide additional polarization observables through angular decay distribution, such as the hadron-side asymmetries in the decay $\Lambda_c^+\to\Lambda^0 \pi^+$ and azimuthal correlations between the two final-state decay planes~\cite{Gutsche:2015mxa}. While the $\Lambda_b$ baryons are not produced at an $e^+ e^-$ B-factory, they account for about $20\%$ of the $b$-hadrons produced at the LHC~\cite{Aaij:2011jp}, making the experimental study of these decays very promising in the near future. It would be, therefore, very interesting to make a comprehensive analysis of the scalar leptoquark effect in these baryonic decays, which will be presented in a forthcoming work~\cite{Lambdab2Lambdac}.

\section*{Acknowledgements}

The work is supported by the National Natural Science Foundation of China (NSFC) under contract Nos. 11005032, 11225523, 11221504 and 11435003. XL is also supported by the Scientific Research Foundation for the Returned Overseas Chinese Scholars, State Education Ministry, and by the self-determined research funds of CCNU from the colleges' basic research and operation of MOE~(CCNU15A02037). XZ is supported by the CCNU-QLPL Innovation Fund (QLPL2015P01).

\begin{appendix}

\section{Analytic formulae of $B_c^-\to\gamma\tau^-\bar\nu_\tau$, $\bar B\to D^{(\ast)}\tau\bar\nu_\tau$ and $B\to X_c\tau\bar\nu_\tau$}
\label{sec:analytic results}

In this appendix, we give all the relevant formulae used to calculate the observables in $B_c^-\to\gamma\tau^-\bar\nu_\tau$, $\bar B\to D^{(\ast)}\tau\bar\nu_\tau$ and $B\to X_c\tau\bar\nu_\tau$ decays.

\subsection{The radiative leptonic decay $B_c^-\to\gamma\tau^-\bar\nu_\tau$}

\begin{figure}[hb]
\centering
  \subfigure[]{\includegraphics[width=0.31\textwidth]{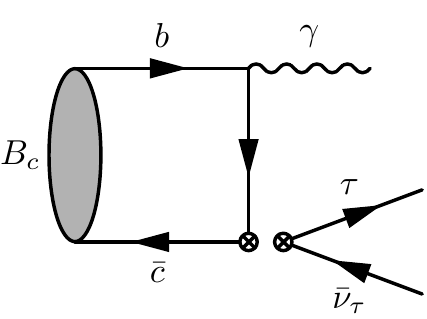}}~
  \subfigure[]{\includegraphics[width=0.31\textwidth]{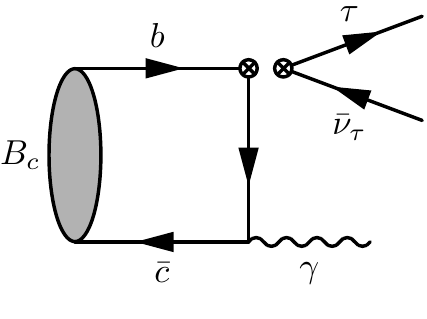}}~
  \subfigure[]{\includegraphics[width=0.31\textwidth]{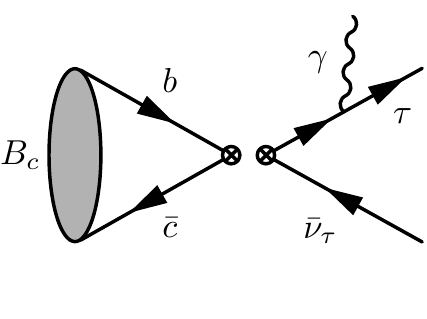}}~
   \caption{\label{fig:Feyn} The leading-order Feynman diagrams for $B_c\to\gamma\tau\bar\nu_\tau$, starting with the effective Hamiltonian Eq.~\eqref{eq:total Hamiltonian}.}
\end{figure}

Starting from the effective Hamiltonian Eq.~\eqref{eq:total Hamiltonian}, we find that there are only three tree-level Feynman diagrams contributing to $B_c^-\to\gamma\tau\bar\nu_\tau$, which are shown in Fig.~\ref{fig:Feyn}.

To calculate these Feynman diagrams, we adopt the peaking approximation for the $B_c$-meson wave functions, $\phi_{B_c}(x_c)=\delta(x_c-\frac{m_c}{m_{B_c}})$, with $x_c$ being the momentum fraction of the $c$ quark~\cite{Brodsky:1985cr,Korner:1991zx,Choi:1992ex,Carlson:1992hz,Du:1996ss}. The spinor part of the $B_c$-meson projector is given by~\cite{Grozin:1996pq,Beneke:2000ry}
\begin{equation}
-\frac{i}{4}f_{B_c}(p\!\!\!\slash_{B_c}+m_{B_c})\gamma_5\,.
\end{equation}
Then we get the amplitudes for Figs.~\ref{fig:Feyn}(a), \ref{fig:Feyn}(b) and \ref{fig:Feyn}(c) containing both the SM and the scalar leptoquark contributions
\begin{align}
\mathcal A_{(a)}=&-\frac{i4G_F}{\sqrt2}V_{cb} (-\frac{ie}{3})(-\frac{i}{4}f_{B_c})\frac{i}{\bar{p}_b^2-m_b^2}\Big\{{\rm Tr} \left[C_V\gamma_\mu P_L(\bar{p}_b\!\!\!\!\!\slash\,+m_b)\gamma_\rho(p_{B_c}\!\!\!\!\!\!\!\!\slash\;\;+m_{B_c})
\gamma_5\right]{\epsilon^\rho}^\ast \bar\tau\gamma^\mu P_L\nu_\tau\nonumber\\
&\hspace{4.2cm}+{\rm Tr} \left[C_S P_L(\bar{p}_b\!\!\!\!\!\slash\,+m_b)\gamma_\rho(p_{B_c}\!\!\!\!\!\!\!\!\slash\;\;+m_{B_c})
\gamma_5\right]{\epsilon^\rho}^\ast \bar\tau P_L\nu_\tau \nonumber\\
&\hspace{4.2cm}+{\rm Tr} \left[C_T\sigma_{\mu\nu} P_L(\bar{p}_b\!\!\!\!\!\slash\,+m_b)\gamma_\rho(p_{B_c}\!\!\!\!\!\!\!\!\slash\;\;+m_{B_c})
\gamma_5\right]{\epsilon^\rho}^\ast \bar\tau \sigma^{\mu\nu}P_L\nu_\tau\Big\}\,,\\[0.2cm]
\mathcal A_{(b)}=&-\frac{i4G_F}{\sqrt2}V_{cb} (\frac{2ie}{3})(-\frac{i}{4}f_{B_c})\frac{i}{\bar{p}_c^2-m_c^2}\Big\{{\rm Tr}\left[C_V\gamma_\rho (\bar{p}_c\!\!\!\!\!\slash\,+m_c)\gamma_\mu P_L(p_{B_c}\!\!\!\!\!\!\!\!\slash\;\;
+m_{B_c})\gamma_5\right]{\epsilon^\rho}^\ast \bar\tau\gamma^\mu P_L\nu_\tau\nonumber\\
&\hspace{4.2cm}+{\rm Tr}\left[C_S\gamma_\rho (\bar{p}_c\!\!\!\!\!\slash\,+m_c)P_L(p_{B_c}\!\!\!\!\!\!\!\!\slash\;\;
+m_{B_c})\gamma_5\right]{\epsilon^\rho}^\ast \bar\tau P_L\nu_\tau \nonumber\\
&\hspace{4.2cm}+{\rm Tr} \left[C_T\gamma_\rho (\bar{p}_c\!\!\!\!\!\slash\,+m_c)\sigma_{\mu\nu}P_L(p_{B_c}\!\!\!\!\!\!\!\!\slash\;\;
+m_{B_c})\gamma_5\right]{\epsilon^\rho}^\ast\bar\tau\sigma^{\mu\nu}P_L\nu_\tau\Big\}
\,,\\[0.2cm]
\mathcal A_{(c)}=&-\frac{i4G_F}{\sqrt2}V_{cb}(-ie)\frac{-if_{B_c}}{2}\frac{i}
{\bar{p}_\tau^2-m_\tau^2}\Big[C_Vp_{B_c}^\mu\bar\tau
\gamma_\rho(\bar{p}\!\!\!\slash_\tau+m_\tau)\gamma_\mu P_L\nu_\tau{\epsilon^\rho}^\ast \nonumber\\
&\hspace{4.2cm} -C_S m_{B_c}\bar\tau
\gamma_\rho(\bar{p}\!\!\!\slash_\tau+m_\tau)P_L\nu_\tau{\epsilon^\rho}^\ast\Big]\,,
\end{align}
where $C_{V,S,T}$ are the Wilson coefficients of the corresponding operators at the scale $\mu=m_b$, and $\bar{p}_{b}=p_b-k$, $\bar{p}_{c}=k-p_c$ and $\bar{p}_{\tau}=p_{\tau}+k$ are, respectively, the momentum of the $b$, $c$ and $\tau$ propagators, with $p_b=\frac{m_b}{m_{B_c}}p_{B_c}$, $p_c=\frac{m_c}{m_{B_c}}p_{B_c}$ and $k$ the momentum of the photon. The photon energy dependence of the differential branching ratio is given by
\begin{align}
{\rm d}\mathcal B(B_c^-\to\gamma\tau^-\bar\nu_\tau)/{\rm d}E_\gamma=\int{\rm d}E_{\nu_\tau}\frac{\tau_{B_c}}{2m_{B_c}}\left|\mathcal A_{(a)}+\mathcal A_{(b)}+\mathcal A_{(c)}\right|^2{\rm d}\Phi_3\,,
\end{align}
where ${\rm d}\Phi_3=\frac{1}{32\pi^3}{\rm d}E_\gamma{\rm d}E_{\nu_\tau}$ is the three-body phase space. Integrating over the photon energy $E_\gamma$, one can then obtain the total branching ratio.

\subsection{Exclusive semi-leptonic decays $\bar B\to D^{(\ast)}\tau\bar\nu_\tau$}

For the exclusive semi-leptonic $\bar B\to D\tau\bar\nu_\tau$ and $\bar B\to D^{\ast}\tau\bar\nu_\tau$ decays, we follow the helicity amplitude formalism that is commonly used in the literatures~\cite{Fajfer:2012vx,Tanaka:2012nw,Sakaki:2013bfa,Becirevic:2016hea,Dorsner:2016wpm}. For simplicity, we list below the relevant formulae without any detailed derivations.
\begin{description}
\item[(1)] $\bar B\to D\tau\bar\nu_\tau$\\[-0.8cm]
\begin{itemize}
\item The differential decay rates
\begin{align}\label{eq:dBrD}
{{\rm d}\Gamma^{\lambda_\tau=1/2}\over{\rm d}q^2}=&{G_F^2 |V_{cb}|^2 \over 192\pi^3 m_B^3} q^2 \sqrt{\lambda(q^2)} \left( 1 - {m_\tau^2 \over q^2} \right)^2 \,\biggl\{\biggr.{1 \over 2}|C_V|^2 {m_\tau^2 \over q^2}\left( H_0^{s\,2}+3H_t^{s\,2}\right)+{3\over2}|C_S|^2\nonumber \\
& H^{s\,2}+8|C_T|^2\,H_{+-}^{s\,2}+3\Re[C_VC_S^\ast]{m_\tau\over\sqrt{q^2}}\,H^sH_t^s-
4\Re[C_VC_T^\ast]{m_\tau\over\sqrt{q^2}}H_{+-}^s H_0^s\biggl.\biggr\}\,,\nonumber\\[0.2cm]
{{\rm d}\Gamma^{\lambda_\tau=-1/2}\over{\rm d}q^2}=&{G_F^2|V_{cb}|^2\over 192\pi^3m_B^3}q^2 \sqrt{\lambda(q^2)}\left(1-{m_\tau^2\over q^2}\right)^2\,\biggl\{\biggr.|C_V|^2H_0^{s\,2}+16|C_T|^2{m_\tau^2\over q^2}H_{+-}^{s\,2}\nonumber\\
&-8\Re[C_VC_T^\ast]{m_\tau\over\sqrt{q^2}}H_{+-}^s H_0^s\biggl.\biggr\} \,,\nonumber\\[0.2cm]
{{\rm d}\Gamma\over{\rm d}q^2}=&{{\rm d}\Gamma^{\lambda_\tau=1/2}\over{\rm d}q^2}+{{\rm d}\Gamma^{\lambda_\tau=-1/2}\over{\rm d}q^2}\,,
\end{align}
where $\lambda^{(*)}(q^2)=\lambda(m_B^2,m_{D^{(*)}}^2,q^2)$ with $\lambda(a,b,c)=a^2+b^2+c^2-2(a b + b c + c a)$, $H$s are the hadronic amplitudes given in Appendix~\ref{sec:amplitudes}, and all the Wilson coefficients $C_i~(i=V,S,T)$ are evaluated at the scale $\mu_b=m_b$.

\item The $q^2$ dependent ratio
\begin{equation}
R_D(q^2)=\frac{{\rm d}\Gamma(\bar B\to D\tau\bar{\nu}_{\tau})/{\rm d}q^2}{{\rm d}\Gamma(\bar B\to D\ell\bar{\nu}_\ell)/{\rm d}q^2}\,,
\end{equation}
where $\ell$ denotes the light lepton~($e$ or $\mu$).

\item The longitudinal polarization of $\tau$
\begin{equation}
P_{\tau}^D(q^2)=\frac{{\rm d}\Gamma^{\lambda_{\tau}=1/2}/{\rm d}q^2-
    {\rm d}\Gamma^{\lambda_{\tau}=-1/2}/{\rm d}q^2}{{\rm d}\Gamma^{\lambda_{\tau}=1/2}/{\rm d}q^2+
    {\rm d}\Gamma^{\lambda_{\tau}=-1/2}/{\rm d}q^2}\,.
\end{equation}

\item The lepton forward-backward asymmetry
\begin{equation}\label{eq:AFB}
A_{\rm FB}(q^2) = \frac{\int_{0}^{1} {\rm d}\cos\theta({\rm d}^2\Gamma/{\rm d}q^2{\rm d}\cos\theta)-\int_{-1}^{0}{\rm d}\cos\theta({\rm d}^2\Gamma/{\rm d}q^2{\rm d} \cos\theta )}{{\rm d}\Gamma/{\rm d}q^2}\,,
\end{equation}
where $\theta$ is the angle between the three-momentum of $\tau$ and that of the $D$ meson in the $\tau$-$\bar{\nu}_{\tau}$ center-of-mass frame. Writing the double-differential decay rates as~\cite{Sakaki:2013bfa}
\begin{equation}
{{\rm d}^2\Gamma\over{\rm d}q^2{\rm d}\cos\theta}= a_\theta(q^2)+b_\theta^{D^{(\ast)}}(q^2)\cos\theta+c_\theta(q^2)\cos^2\theta\,,
\end{equation}
one can then see clearly that the coefficient $b_\theta$ determines the lepton forward-backward asymmetry, with
\begin{align}
A_{\rm FB}^D(q^2)= b_\theta^D(q^2)=&{G_F^2|V_{cb}|^2\over128\pi^3m_B^3}q^2\sqrt{\lambda(q^2)} \left(1-{m_\tau^2\over q^2}\right)^2\,\biggl\{\biggr.|C_V|^2{m_\tau^2\over q^2}H_0^sH_t^s+\Re[C_VC_S^\ast] \nonumber\\
& \hspace{-1.5cm} {m_\tau\over\sqrt{q^2}}H^sH_0^s -4\Re[C_VC_T^\ast]{m_\tau\over\sqrt{q^2}}H_{+-}^sH_t^s
-4\Re[C_SC_T^\ast]H_{+-}^sH^s\biggl.\biggr\}\,,
\end{align}
and
\begin{align}
A_{\rm FB}^{D^\ast}(q^2)= b_\theta^{D^\ast}(q^2)=&{G_F^2|V_{cb}|^2\over128\pi^3m_B^3}q^2 \sqrt{\lambda^\ast(q^2)}\left(1-{m_\tau^2\over q^2}\right)^2\,\biggl\{\biggr.|C_V|^2\Big[{1\over2}\left(H_+^{+\,2}-H_-^{-\,2}\right)
\nonumber\\
& \hspace{-1.5cm} +{m_\tau^2\over q^2}H_0^0H_t^0\Big]+8|C_T|^2{m_\tau^2\over q^2}\cdot\left( H_{+t}^{0\,2}-H_{-t}^{0\,2}\right)-\Re[C_VC_S^\ast] {m_\tau \over \sqrt{q^2}}\nonumber\\
& \hspace{-1.5cm}  H^0H_0^0\!-\!4\Re[C_VC_T^\ast]{m_\tau\over\sqrt{q^2}}\!\left(H_{+-}^0H_t^0\!+\!H_{+t}^0
H_+^+\!-\!H_{-t}^0H_-^- \right)\nonumber\\
& \hspace{-1.5cm} +4\Re[C_SC_T^\ast]H_{+-}^0H^0\biggl.\biggr\}\,.
\end{align}

\end{itemize}

\item[(2)] $\bar B\to D^{\ast}\tau\bar\nu_\tau$\\[-0.8cm]
\begin{itemize}
\item The differential decay rates\\[-1.0cm]
\end{itemize}
\begin{align}\label{eq:dBrDst}
{{\rm d}\Gamma^{\lambda_\tau=1/2}\over{\rm d}q^2}=&{G_F^2|V_{cb}|^2\over 192\pi^3 m_B^3}q^2 \sqrt{\lambda^\ast(q^2)}\left(1-{m_\tau^2\over q^2}\right)^2\,\biggl\{\biggr.{1\over2}|C_V|^2{m_\tau^2\over q^2}\left(H_+^{+\,2}+H_-^{-\,2}+H_0^{0\,2}+H_t^{0\,2}\right)\nonumber\\[0.1cm]
&+{3\over2}|C_S|^2H^{0\,2}+8|C_T|^2\left(H_{+t}^{+\,2}+H_{-t}^{-\,2}+H_{+-}^{0\,2}\right)
-3\Re[C_VC_S^\ast]{m_\tau\over\sqrt{q^2}}H^0H_t^0\nonumber\\
&-4\Re[C_VC_T^\ast]{m_\tau\over\sqrt{q^2}}\left(H_{+-}^0H_0^0+H_{+t}^+H_+^++H_{-t}^-H_-^- \right)\biggl.\biggr\}\,,\nonumber \\[0.2cm]
{{\rm d}\Gamma^{\lambda_\tau=-1/2}\over {\rm d}q^2}=&{G_F^2|V_{cb}|^2\over 192\pi^3 m_B^3}q^2 \sqrt{\lambda^\ast(q^2)}\left(1-{m_\tau^2\over q^2}\right)^2\,\biggl\{\biggr.|C_V|^2\left(H_+^{+\,2}+H_-^{-\,2}+H_0^{0\,2}\right)
+16|C_T|^2{m_\tau^2\over q^2}\nonumber\\[0.1cm]
& \left( H_{+t}^{+\,2}+H_{-t}^{-\,2}+H_{+-}^{0\,2}\right)-8\Re[C_VC_T^\ast]{m_\tau \over \sqrt{q^2}}\left( H_{+-}^0H_0^0+ H_{+t}^+H_+^++H_{-t}^-H_-^-\right)\biggl.\biggr\}\,,\nonumber\\[0.2cm]
{{\rm d}\Gamma^{\lambda_{D^\ast}=\pm1}\over{\rm d}q^2}=&{G_F^2|V_{cb}|^2\over192\pi^3 m_B^3}q^2\sqrt{\lambda^\ast(q^2)}\left(1-{m_\tau^2\over q^2}\right)^2\,\biggl\{ \biggr.\left(1+{m_\tau^2\over2q^2}\right)\bigl(\bigr.|C_V|^2 H_\pm^{\pm\,2}+8|C_T|^2\nonumber\\[0.1cm]
& \left(1+{2m_\tau^2\over q^2}\right) H_{\pm t}^{\pm\,2}\mp12\Re[C_VC_T^\ast]{m_\tau\over\sqrt{q^2}}H_{\pm t}^\pm H_\pm^\pm\biggl.\biggr\}\,,\nonumber\\[0.2cm]
{{\rm d}\Gamma^{\lambda_{D^\ast}=0}\over{\rm d}q^2}=&{G_F^2 |V_{cb}|^2 \over 192\pi^3 m_B^3}q^2\sqrt{\lambda^\ast(q^2)}\left(1-{m_\tau^2\over q^2}\right)^2 \,\biggl\{\biggr.|C_V|^2\left[\left(1+{m_\tau^2\over2q^2}\right)H_0^{0\,2}
+{3\over2}{m_\tau^2\over q^2}H_t^{0\,2}\right]\nonumber\\[0.1cm]
&+{3\over 2}|C_S|^2H^{0\,2}+8|C_T|^2\left(1+{2m_\tau^2\over q^2}\right)H_{+-}^{0\,2}-3\Re[C_VC_S^\ast]{m_\tau\over\sqrt{q^2}}H^0H_t^0 \nonumber\\[0.1cm]
&-12\Re[C_VC_T^\ast]{m_\tau \over\sqrt{q^2}}H_{+-}^0H_0^0\biggl.\biggr\}\,.
\end{align}
Besides the observables similar to that defined in $B^-\to D\tau^-\bar\nu_\tau$, there are another two observables in this process, i.e., the longitudinal and transverse polarizations of the $D^\ast$ meson defined, respectively, by
\begin{align}
P_L^{D^{\ast}}(q^2) =& \frac{{\rm d}\Gamma^{\lambda_{D^\ast}=0}/{\rm d}q^2}{{\rm d}\Gamma^{\lambda_{D^\ast}=0}/{\rm d}q^2+{\rm d}\Gamma^{\lambda_{D^\ast}=1}/{\rm d}q^2+{\rm d}\Gamma^{\lambda_{D^\ast}=-1}/{\rm d}q^2}\,, \\[0.2cm]
P_T^{D^{\ast}}(q^2) =& 1-P_L^{D^{\ast}}(q^2)\,.
\end{align}

\end{description}

\subsection{Inclusive semi-leptonic decay $B\to X_c\tau\bar\nu_\tau$}

In the heavy-quark limit $m_b\gg\Lambda_{\rm QCD}$, the inclusive semi-leptonic decay rate is equivalent to the perturbative quark-level $b$ decay rate~\cite{Manohar:2000dt,Falk:1994gw,Chay:1990da,Shifman:1984wx}. This makes it possible to get the inclusive decay rate by calculating directly the rate for the quark-level process $b\to c\tau \bar\nu_{\tau}$, with the result given by
\begin{align}\label{eq:inclusive width}
\frac{{\rm d}\Gamma}{{\rm d}\hat q^2}=&\frac{G_F^2V_{cb}^2m_b^5}{192\pi^3}\!\left(1\!-\!\frac{x_\tau}{\hat q^2}\right)^2\! \!\lambda^{1/2}(1,x_c,\hat q^2)\bigg[3|C_S|^2(1\!+\!x_c\!-\!\hat q^2)\hat q^2\!+\!6\Re{[C_VC_S^{\ast}]}x_c^{1/2}x_\tau^{1/2}(1\!-\!x_c\!+\!\hat q^2)\nonumber\\
&+16|C_T|^2(2x_c^2\!-\!x_c(\hat q^2\!+\!4)\!-\!\hat q^4\!-\!\hat q^2\!+\!2)\left(1\!+\!\frac{2x_\tau}{\hat q^2}\right)\!-\!72\Re{[C_VC_T^{\ast}]}x_c^{1/2} x_\tau^{1/2}(1\!-\!x_c\!+\!\hat q^2)\nonumber\\
&+2|C_V|^2\Big((1+x_c-x_\tau)\hat q^2+(1-x_c)^2-x_\tau(1+x_c)+2(1-x_c)^2\frac{x_\tau}{\hat q^2}-2\hat q^4\Big)\bigg]\,,
\end{align}
where $x_i=\frac{m_i^2}{m_b^2}$, and $\hat q^2=\frac{q^2}{m_b^2}$ with $\hat q^2$ varying from $x_\tau$ to $(1-\sqrt{x_c})^2$. The $\tau$-energy spectrum of this process is given by
\begin{align}\label{eq:tau-energy-spectrum}
\frac{{\rm d}\Gamma}{{\rm d}y}=&\frac{G_F^2V_{cb}^2m_b^5}{192\pi^3} \frac{\sqrt{y^2-4x_\tau}(1+x_\tau-x_c-y)^2}
{2(1+x_\tau-y)^3}\bigg[\Big(4|C_V|^2+|C_S|^2\Big)\Big(2y^3-(x_c+5x_\tau+5)y^2\nonumber\\
&+\left(3x_c(x_\tau+1)+(3x_\tau^2+10x_\tau+3)\right)y-4x_\tau(2x_c+x_\tau+1)\Big)+16|C_T|^2\Big(14y^3\nonumber\\
&-(x_c+29x_\tau+29)y^2+\left(3x_c(x_\tau+1)+15x_\tau^2+34x_\tau+15\right)y
-4x_\tau(2x_c+x_\tau+1)\Big)\nonumber\\
&-12\Big(\Re[C_VC_S^\ast]-12\Re[C_VC_T^\ast]\Big)\sqrt{x_c}\sqrt{x_\tau}(1+x_\tau-y)(y-2)+8\Re[C_SC_T^\ast]\Big(4y^3\nonumber\\
&+(x_c-7 x_\tau-7)y^2-\left(3x_c(x_\tau+1)-3 x_\tau^2-2x_\tau-3\right)y+4x_\tau(2x_c+x_\tau+1)\Big)\bigg]\,,
\end{align}
where $y=\frac{2 E_{\tau}}{m_b}$, with $y$ varying from $2\sqrt{x_\tau}$ to $1+x_\tau-x_c$.

\section{Hadronic amplitudes in $\bar B\to D^{(\ast)}\tau\bar\nu_\tau$ decays}
\label{sec:amplitudes}

The hadronic amplitudes in $\bar B\to D^{(\ast)}\tau\bar\nu_\tau$ decays are given, respectively,  as~\cite{Fajfer:2012vx,Tanaka:2012nw,Sakaki:2013bfa,Dorsner:2016wpm}
\begin{itemize}
\item $\bar B\to D\tau\bar \nu_\tau$\\[-1.0cm]
\end{itemize}
   \begin{align}
         H_0^s(q^2)=& \sqrt{\lambda_D(q^2) \over q^2} F_1(q^2)\,, &  H_t^s(q^2)=& {m_B^2-m_D^2 \over \sqrt{q^2}} F_0(q^2) \,,\nonumber \\[0.2cm]
         H^s(q^2)=&{m_B^2-m_D^2 \over m_b-m_c} F_0(q^2) \,,&
         H^s_{+-}(q^2)=& H_{0t}^s(q^2) = -{\sqrt{\lambda_D(q^2)} \over m_B+m_D} F_T(q^2) \,.
   \end{align}

\begin{itemize}
\item $\bar B\to D^\ast\tau\bar \nu_\tau$\\[-1.0cm]
\end{itemize}
   \begin{align}
         H_\pm^\pm(q^2) =& (m_B+m_{D^\ast}) A_1(q^2) \mp { \sqrt{\lambda_{D^\ast}(q^2)} \over m_B+m_{D^\ast} } V(q^2) \,,\nonumber \\[0.2cm]
         H_0^0(q^2)=& { m_B+m_{D^\ast} \over 2m_{D^\ast}\sqrt{q^2} } \left[ -(m_B^2-m_{D^\ast}^2-q^2) A_1(q^2)+{ \lambda_{D^\ast}(q^2) \over (m_B+m_{D^\ast})^2 } A_2(q^2) \right] \,,\nonumber \\[0.2cm]
         H_t^0(q^2)=&-\sqrt{ \lambda_{D^\ast}(q^2) \over q^2 } A_0(q^2) \,,\quad
         H^0(q^2)={ \sqrt{\lambda_{D^\ast}(q^2)} \over m_b+m_c } A_0(q^2) \,,\nonumber\\[0.2cm]
         H_{\pm t}^\pm(q^2) =& { 1 \over \sqrt{q^2} } \left[ (m_B^2-m_{D^\ast}^2) T_2(q^2) \pm\sqrt{\lambda_{D^\ast}(q^2)} T_1(q^2) \right] \,,\nonumber\\[0.2cm]
         H_{+-}^0(q^2)=&H_{0t}^0(q^2)={1 \over 2m_{D^\ast} } \left[ -(m_B^2+3m_{D^\ast}^2-q^2) T_2(q^2)+ { \lambda_{D^\ast}(q^2) \over m_B^2-m_{D^\ast}^2 } T_3(q^2) \right] \,,
   \end{align}
where all the $B\to D$ and $B\to D^{\ast}$ form factors are taken from Refs.~\cite{Sakaki:2013bfa,Caprini:1997mu}, except for $S_1(w)$, for which we use the form given by Eq.~(A6) in Ref.~\cite{Sakaki:2012ft}.

\end{appendix}

\bibliographystyle{JHEP}



\begingroup\raggedright

\endgroup

\end{document}